\documentclass[lettersize,journal]{IEEEtran}

\usepackage{array}
\usepackage[caption=false,font=normalsize,labelfont=sf,textfont=sf]{subfig}
\usepackage{textcomp}
\usepackage{stfloats}
\usepackage{url}
\usepackage{hyperref}
\usepackage[normalem]{ulem}
\useunder{\uline}{\ul}{}
\usepackage{verbatim}
\usepackage{graphicx}
\usepackage{amsmath,amssymb}
\usepackage{wrapfig}
\usepackage[ruled]{algorithm2e} 

\usepackage{overpic}
\usepackage{booktabs} 
\usepackage{multirow}

\usepackage{scalerel}
\usepackage{tikz}
\usetikzlibrary{svg.path}

\definecolor{orcidlogocol}{HTML}{A6CE39}
\tikzset{
  orcidlogo/.pic={
    \fill[orcidlogocol] svg{M256,128c0,70.7-57.3,128-128,128C57.3,256,0,198.7,0,128C0,57.3,57.3,0,128,0C198.7,0,256,57.3,256,128z};
    \fill[white] svg{M86.3,186.2H70.9V79.1h15.4v48.4V186.2z}
                 svg{M108.9,79.1h41.6c39.6,0,57,28.3,57,53.6c0,27.5-21.5,53.6-56.8,53.6h-41.8V79.1z M124.3,172.4h24.5c34.9,0,42.9-26.5,42.9-39.7c0-21.5-13.7-39.7-43.7-39.7h-23.7V172.4z}
                 svg{M88.7,56.8c0,5.5-4.5,10.1-10.1,10.1c-5.6,0-10.1-4.6-10.1-10.1c0-5.6,4.5-10.1,10.1-10.1C84.2,46.7,88.7,51.3,88.7,56.8z};
  }
}

\newcommand\orcidicon[1]{\href{https://orcid.org/#1}{\mbox{\scalerel*{
\begin{tikzpicture}[yscale=-1,transform shape]
\pic{orcidlogo};
\end{tikzpicture}
}{|}}}}
\AtBeginDocument{%
  }
\usepackage{color}
\definecolor{AAA}{rgb}{1.0, 0.13, 0.32}
\definecolor{BBB}{rgb}{0.2, 0.1, 1}
\definecolor{CCC}{rgb}{0.0, 0.0, 0.0}

\newcommand{\CG}[1]{{\color{CCC}#1}}

\definecolor{bittersweet}{rgb}{1.0, 0, 0.5}

\usepackage{cite}
\begin{document}

\title{Efficient Nearest Neighbor Search Using Dynamic Programming}

\author{\href{https://pengfei.me}{Pengfei Wang} \orcidicon{0000-0002-2079-275X},
Jiantao Song \orcidicon{0009-0007-0907-3731}, 
Shiqing Xin (Corresponding author) \orcidicon{0000-0001-8452-8723}, 
Shuangmin Chen \orcidicon{0000-0002-0835-3316},
Changhe Tu \orcidicon{0000-0002-1231-3392},
Wenping Wang \orcidicon{0000-0002-2284-3952} ~\IEEEmembership{Fellow,~IEEE}, 
Jiaye Wang

\IEEEcompsocitemizethanks{
\IEEEcompsocthanksitem Pengfei Wang, Jiantao Song, Shiqing Xin, and Changhe Tu are with the School of Computer Science and Technology, Shandong University, Qingdao 266237, China (e-mail: \href{mailto:zihanzhao2000@gmail.com}{zihanzhao2000@gmail.com}; \href{pengfei1998@foxmail.com}{pengfei1998@foxmail.com}; \href{mailto:xinshiqing@sdu.edu.cn}{xinshiqing@sdu.edu.cn}; \href{mailto:chtu@sdu.edu.cn}{chtu@sdu.edu.cn}).
\IEEEcompsocthanksitem Minfeng Xu is with School of Computer Science and Technology, Shandong University of Finance and Economics, Jinan 250014, China (e-mail: \href{mailto:mfxu_sdu@163.com}{mfxu\_sdu@163.com}).
\IEEEcompsocthanksitem Shuangmin Chen is with the School of Information and Technology, Qingdao University of Science and Technology, Shandong 266101, China (e-mail: \href{mailto:csmqq@163.com}{csmqq@163.com}).
\IEEEcompsocthanksitem Wenping Wang is with the Department of Computer Science and Engineering, Texas A\&M University, College Station, TX 77843 USA (e-mail:
\href{mailto:wenping@tamu.edu}{wenping@tamu.edu}).
\IEEEcompsocthanksitem *Shiqing Xin is the corresponding author.
}
}




\maketitle
\begin{abstract}
Given a collection of points in $\mathbb{R}^3$,
KD-Tree and R-Tree are well-known nearest neighbor search (NNS) algorithms that rely on space partitioning and spatial indexing techniques. 
However, when the query point is far from the data points or the data points inherently represent a 2-manifold surface, their query performance may degrade. 

To address this, we propose a novel dynamic programming technique that precomputes a Directed Acyclic Graph (DAG) to encode the proximity structure between data points. More specifically, the DAG captures how the proximity structure evolves during the incremental construction of the Voronoi diagram of the data points. 
Experimental results demonstrate that our method achieves a 1-10x speedup.

\CG{
Additionally, our algorithm demonstrates significant practical value across diverse applications. We validated its effectiveness through extensive testing in four key applications: Point-to-Mesh Distance Queries, Iterative Closest Point (ICP) Registration, Density Peak Clustering, and Point-to-Segments Distance Queries. A particularly notable feature of our approach is its unique ability to efficiently identify the nearest neighbor among the first $k$ points in the point cloud—a capability that enables substantial acceleration in low-dimensional applications like Density Peak Clustering. As a natural extension of our incremental construction process, our method can also be readily adapted for farthest point sampling tasks. These experimental results across multiple domains underscore the broad applicability and practical importance of our approach.
}

\end{abstract}


\section{Introduction}
\IEEEPARstart{G}{iven} a target point set $\mathcal{P}=\{p_i\}_{i=1}^n$ and a query point \( q \), the point \( p_i \) is considered the nearest neighbor in $\mathcal{P}$ to \( q \) if \( p_i \) satisfies
\begin{equation}
    \forall j \in [1,n], \|q-p_i\| \leq \|q-p_j\|,
\end{equation}
where \( \| \cdot \| \) denotes the Euclidean distance.
Finding the nearest neighbor is a fundamental operation in various fields, including computer graphics~\cite{Rui2022RFEPS, pearson1901liii, Zong2023P2M}, computer vision~\cite{corso2021classification,liu2007clustering}, and robotics~\cite{quin2017experimental}.
Due to its critical applications across various fields, improving query speed remains an active area of research. 
Traditional nearest-neighbor query methods include KD Trees~\cite{10.1145/361002.361007} and R Trees~\cite{10.1145/602259.602266}, which offer fast query speeds for general point clouds. These methods achieve efficiency by constructing hierarchical structures to partition the space or point cloud, enabling quick localization during queries. 
However, when handling data distributed along 2D manifold surfaces embedded in 3D space (referred to as 2D manifold data), common in computer graphics, or when the query point lies far from the dataset, the efficiency of these traditional algorithms can decline. In extreme cases, the query complexity may degrade to \(O(n)\).

As shown in the inset figure, we consider two point clouds with distinct distributions: one uniformly distributed within a unit cube and the other across the surface of the Earth. Additionally, we generate two sets of query points—one within the bounding box of the input data and the other within 8 times the bounding box. It is evident that the query performance of both KD Trees and R Trees deteriorates significantly under these conditions.
\begin{wrapfigure}{r}{0.25\textwidth}
  \centerline{
  \begin{overpic}
  [width=0.25\textwidth]{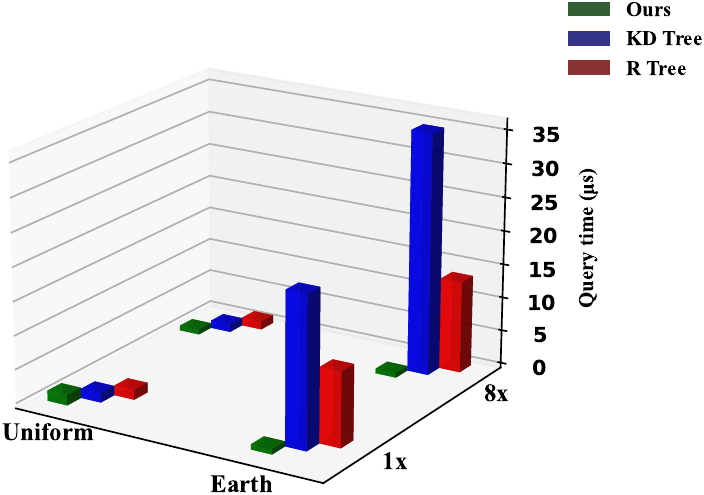}
  \end{overpic}
  } 
  \vspace*{-3mm}
\end{wrapfigure}

In this paper, we present a simple, effective, and novel nearest point query algorithm based on Voronoi / Delaunay.
For 2D manifold data in 3D, our query speed is 1x-10x faster than KD Tree and R Tree. Even for uniformly distributed random point clouds, our query algorithm achieves speeds comparable to KD Tree.
Our main contributions are as follows:
\begin{itemize}
\item We introduced a concise nearest neighbor search algorithm for 2D manifold point cloud data that significantly improves query speed compared to traditional algorithms.
\item If the point cloud possesses an order, our method can facilitate the identification of the nearest point to \( q \) among \( \{p_i\}_{i=1}^k \), for any \( k \), where \( k \leq n \). This approach can be employed to reduce the time complexity of Density Peak Clustering~\cite{rodriguez2014clustering}, thus enhancing its applicability in large-scale scenarios.
\item \CG{We demonstrate our algorithm's practical effectiveness through comprehensive evaluation on four key applications: Point-to-Mesh Distance Queries, Iterative Closest Point Registration, Density Peak Clustering, and Point-to-Segments Distance Queries, showcasing its broad applicability in computational geometry scenarios.}
\end{itemize}


\section{Related Work}

\subsection{Nearest Neighbor Search}
For spatial data, the nearest neighbor problem is typically addressed by building a spatial indexing structure~\cite{6809191}. These techniques can be broadly categorized into two types: those that partition space and those that partition data.

Classic structures for space partitioning include KD Trees and quadtrees. A KD Tree~\cite{10.1145/361002.361007} is a spatial partitioning structure that recursively divides the dataset along one of the \(k\) coordinate system axes, typically alternating between them; for instance, with \(k=2\), splits occur alternately on the x- and y-axes. This splitting process continues until a maximum depth is reached or until fewer than a specified number of points remain in a cell. KD Trees can be extended to include objects beyond points, such as triangles or line segments. Determining the optimal splitting plane can be challenging in these cases; however, for points (or disks) in two dimensions, one can simply sort the points along each dimension and split at the median.
\CG{The construction of a KD Tree can be achieved in $O(n \log n)$ time, and it can find the nearest neighbors in $O(\log n)$ time on average~\cite{10.1145/355744.355745}.}

An octree is another spatial partitioning structure that specifically divides 3D space into eight cubes. This is typically done recursively, with each subdivision occurring when the data within a region meets certain criteria, such as exceeding a maximum number of points or objects. Each internal node in the octree corresponds to a cube region of space, further subdivided into eight equal-sized cubes represented by its children. \CG{The computation cost for locating a leaf node in an octree is $O(depth) \approx O(\log n)$, where depth represents the average depth of the tree~\cite{10.1007/s00138-017-0889-4}. However, nearest neighbor queries typically require additional traversal operations beyond simple leaf node access, potentially increasing the practical query time by either constant factors or additional complexity terms.}

The R Tree~\cite{10.1145/602259.602266} is a classic structure for partitioning data. It is based on B-trees and shares a similar structure: both the number of objects in a leaf node and the number of children in an internal node are bounded by specified minimum and maximum limits, and all leaf nodes are at the same level. 
\CG{
The R*-tree~\cite{rstarTREE} is an enhanced variant that incorporates a combined optimization of area, margin, and overlap of enclosing rectangles. It introduces forced reinsert to dynamically reorganize the structure and prevent unnecessary splits, demonstrating superior performance over existing R-tree variants across different data distributions and query types.
}

Grids and linear search methods are also traditional approaches for nearest-neighbor searches. The grid-based approach improves efficiency by overlaying a regular grid on the entire simulation area and tracking agents within each cell. 
\CG{During query operations, the algorithm examines the cell containing the query point and its adjacent cells. 
However, this approach exhibits significant performance degradation in certain scenarios—particularly when data points are distributed on a spherical surface with the query point at the center. In such cases, the algorithm must traverse multiple cells in all directions, essentially performing a comprehensive dataset scan.
Linear search, conversely, offers a straightforward solution by calculating distances between the query point and each target point, making it suitable only when dealing with very few query operations or small datasets.}

\CG{
Additionally, sketching techniques have been proposed for similarity search acceleration. Mic et al.~\cite{10.1007/978-3-319-98398-1_9} introduced methods for selecting optimal sketches that transform objects into bit-strings. Their sketch construction often relies on reference points, or pivots, to partition the data space. While their approach focuses on Hamming embedding for approximate similarity filtering, our method constructs a `Query Table' based on the incremental insertion of points, which begins the search from a fixed initial point. Our goal is the exact nearest neighbor search in metric spaces through Voronoi diagram construction.}

\subsection{Voronoi Diagram \& Delaunay}
\CG{Given a set of generators $\mathcal{P} = \{p_i\}_{i=1}^n$ in a domain $\Omega$ equipped with a metric function $\mathcal{D}$, the Voronoi diagram partitions $\Omega$ into regions where each generator $p_i$ dominates a cell defined as:}
\begin{equation}
\text{Cell}(p_i) = \{x \in \Omega ~|~ \mathcal{D}(p_i, x) \leq \mathcal{D}(p_j, x), \, \forall j \neq i\}.  
\end{equation}
\CG{While Voronoi diagrams can be generalized across different domains, metrics, and generator types, the most commonly studied variant assumes Euclidean spaces with point generators. As fundamental geometric structures, Voronoi diagrams have found extensive applications in computer graphics~\cite{Zong2023P2M, Rui2022RFEPS, wang2022matfp}.}

\begin{figure}[h]
	\centering
\begin{overpic}
[width=.8\linewidth]{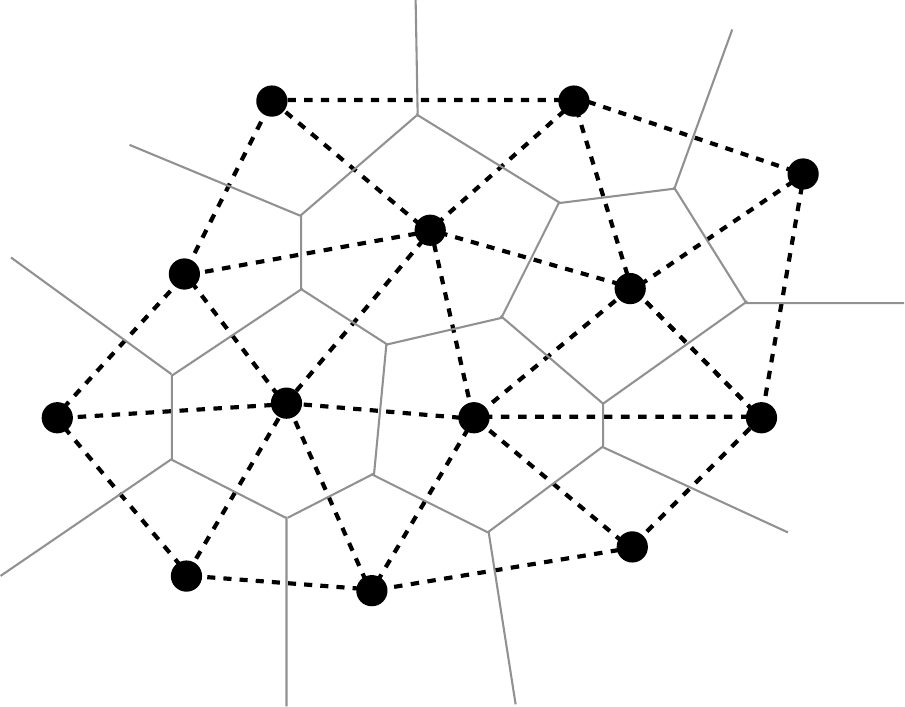}
\end{overpic}
\caption{
\CG{Example of Voronoi diagrams and Delaunay triangulations in 2D. The gray solid lines represent the Voronoi diagram partitioning the plane based on proximity to point generators (black dots), while the black dashed lines show the corresponding Delaunay triangulation that encodes the connectivity relationships between neighboring Voronoi cells.}
}
\label{Fig:voronoiAndDelaunay}
\end{figure}

\CG{The Delaunay triangulation serves as the geometric dual of the Voronoi diagram, encoding the adjacency relationships between Voronoi cells. In 2D, the Delaunay triangulation consists of triangles whose vertices correspond to the Voronoi generators, while in 3D, it comprises tetrahedra with the same vertex correspondence.
Figure~\ref{Fig:voronoiAndDelaunay} illustrates this duality in the 2D setting, where the Delaunay triangulation is depicted with black dashed lines and the Voronoi diagram with gray solid lines.
This organizational structure makes it straightforward to obtain the adjacent sites of any given vertex, with many computational geometry libraries providing corresponding interfaces for such operations.
Currently, Voronoi diagrams are typically computed by first calculating the Delaunay triangulation and then deriving the dual structure. The Bowyer–Watson algorithm~\cite{10.1093/comjnl/24.2.162} is a classic method for computing the Delaunay triangulation. It is based on the empty circle property and works by incrementally inserting points and flipping triangles that do not satisfy this property. Common tools for computing Delaunay triangulations include TetGen~\cite{10.1145/2629697} and CGAL~\cite{cgal:hs-chdt3-24a}, which can achieve an average computational complexity of \(O(n \log n)\) with a worst-case complexity of \(O(n^2)\)~\cite{10.1145/777792.777823}.
}

\CG{
The geometric properties of Voronoi diagrams also enable nearest neighbor queries. By construction, if the closest point to a query location $q$ among generators $\mathbf{P}$ is $p_k$, then $q$ necessarily lies within $\text{Cell}(p_k)$. However, the practical utility of this property is limited by the absence of efficient algorithms for point location within Voronoi cells, presenting an ongoing challenge for large-scale nearest neighbor applications.
}





\begin{figure*}[h]
	\centering
\begin{overpic}
[width=.99\linewidth]{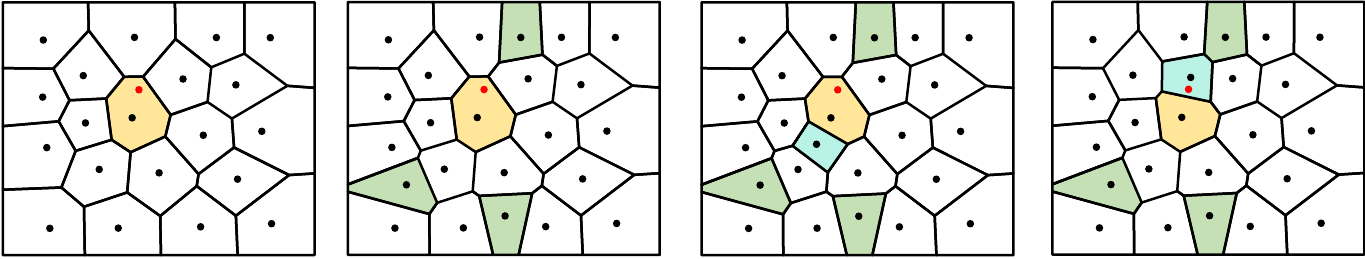}
\put(10, -2){~\textbf{(a)}}
\put(36, -2){~\textbf{(b)}}
\put(61, -2){~\textbf{(c)}}
\put(87, -2){~\textbf{(d)}}
\put(86, 14){\tiny{$p_{24}$}}
\put(59, 9){\tiny{$p_{24}$}}
\put(8, 9){\tiny{$\Phi_{20}(q)$}}
\put(33, 9){\tiny{$\Phi_{23}(q)$}}

\end{overpic}
\vspace{1mm}
\caption{
(a) The Voronoi diagram of $\{p_i\}_{i=1}^{20}$, where the query point $q$ (colored in red) belongs to the Voronoi cell colored in yellow.
(b) If the addition of the three sites (whose cells are colored in green) does not change the yellow cell,
$\Phi_m(q)$ remains unchanged when $m$ is increased from 20 to 23.
\CG{(c)} When the 24-th site $p_{24}$ is added (see the cyan cell), the yellow cell diminishes; yet $\Phi_{24}(q)=\Phi_{23}(q)$ if $q$ is nearer to the old site.
(d) $\Phi_{24}(q)=p_{24}$ if $q$ is nearer to the newly added site $p_{24}$.
}
\label{FIG:dp}
	\vspace{-3mm}
\end{figure*}

\section{Methodology}
\label{Sec:Methodology}
In this section, we introduce the key concept underlying the algorithms, beginning with a solution of linear search.


We assume that the given points are numbered from 1 to $n$. Given a query point $q$, we use $\Phi_m(q)$ to denote the closest point in the subset $\{p_i\}_{i=1}^m$, where $m \leq n$. This leads to the following state transfer equation:
\begin{equation}
\hspace{-3mm}
    \Phi_m(q) = \begin{cases}
        p_1 & \text{if } m = 1, \\
        \Phi_{m-1}(q) & \text{if } m \neq 1 \land \|q - \Phi_{m-1}(q)\| \leq \|q - p_m\|, \\
        p_m & \text{if } m \neq 1 \land \|q - \Phi_{m-1}(q)\| > \|q - p_m\|,
    \end{cases}
\end{equation}
which can be understood as a dynamic programming problem, where $\Phi_n(q)$ is the actual nearest point to be extracted. 
Obviously, a na\"{i}ve implementation of the query operation costs $O(n)$ time. In the following, we consider how to boost this implementation.

Recall that Voronoi diagrams~\cite{Voronoi+1908+97+102} precisely characterize how the space of interest is divided by the given data points. A point $q$ is nearest to $p_i$ if and only if $q$ is located in $p_i$'s cell. However, Voronoi diagrams are not well-suited for NNS due to their lack of hierarchical representation. 
Despite this, during the incremental construction process of Voronoi Diagram, the area dominated by a site either remains unchanged or shrinks, encoding the history of how the proximity structure changes when a new point is added. Let the input point cloud be $\mathcal{P} = \{p_i\}_{i=1}^n$. 
We illustrate our construction process by considering any moment of constructing the Voronoi diagram through incremental point insertion, such as \(\mathcal{V}_m\), which is constructed from the set \(\{p_i\}_{i=1}^m\), where \(m \leq n\).

Figure~\ref{FIG:dp}(a) shows the Voronoi diagram $\mathcal{V}_{20}$ determined by 20 sites, with one of the cells highlighted in yellow. By adding three additional sites (whose cells are highlighted in green), we obtain a new Voronoi diagram $\mathcal{V}_{23}$; see Figure~\ref{FIG:dp}(b). It can be seen that the addition of the three green sites does not change the yellow cell. Therefore, if $q$ is located in the yellow cell, then $\Phi_{23}(q) = \Phi_{20}(q)$, implying that the nearest site to $q$ remains unchanged even when the three green sites are added. 
Only when a newly added site changes the yellow cell may the nearest site differ. Specifically, $\Phi_{24}(q) = \Phi_{23}(q)$ if $q$ is nearer to the old site (see Figure~\ref{FIG:dp}(c)), and $\Phi_{24}(q) = p_{24}$ if $q$ is nearer to the new site $p_{24}$ (see Figure~\ref{FIG:dp}(d)). 
To summarize, by using $p_{m+1} \nvdash \text{Cell}(\Phi_{m}(q);\mathcal{V}_m)$ to represent that the addition of $p_{m+1}$ does not diminish the cell of $\Phi_{m}(q)$ in $\mathcal{V}_m$, we have:
\begin{equation}
    p_{m+1} \nvdash \text{Cell}(\Phi_{m}(q);\mathcal{V}_m) \quad \mapsto \quad \Phi_{m+1}(q) = \Phi_{m}(q),
\end{equation}
or 
\begin{align}
     p_{m+j} \nvdash \text{Cell}&(\Phi_{m}(q);\mathcal{V}_{m+j-1}), \ \forall 1 \leq j \leq k \quad\nonumber\\
      & \quad \quad \quad\downarrow\quad\\
       \mapsto \quad \Phi_{m+k}(q) &= \Phi_{m+k-1}(q) = \cdots = \Phi_{m}(q).\nonumber
\end{align}

\section{Algorithm}
\subsection{Query Table Construction}
\label{sec:queryTableConstruction}
Based on Section~\ref{Sec:Methodology}, we can design a data structure called the Query Table, denoted as $\mathcal{L}=\{\mathcal{L}_i\}_{i=1}^{n}$, to store the necessary information for the query phase.
Specifically, we assign each point $p_i$ a corresponding vector $\mathcal{L}_i$, called Query List.
During the incremental construction of the Voronoi diagram, when \( p_j \) is inserted, if the condition \( p_j \vdash \text{Cell}(p_i;\mathcal{V}_{j-1}) \) is satisfied, we push \( p_j \) to the end of \( \mathcal{L}_i \).

\begin{figure*}[t]
	\centering
\begin{overpic}
 [width=.99\linewidth]{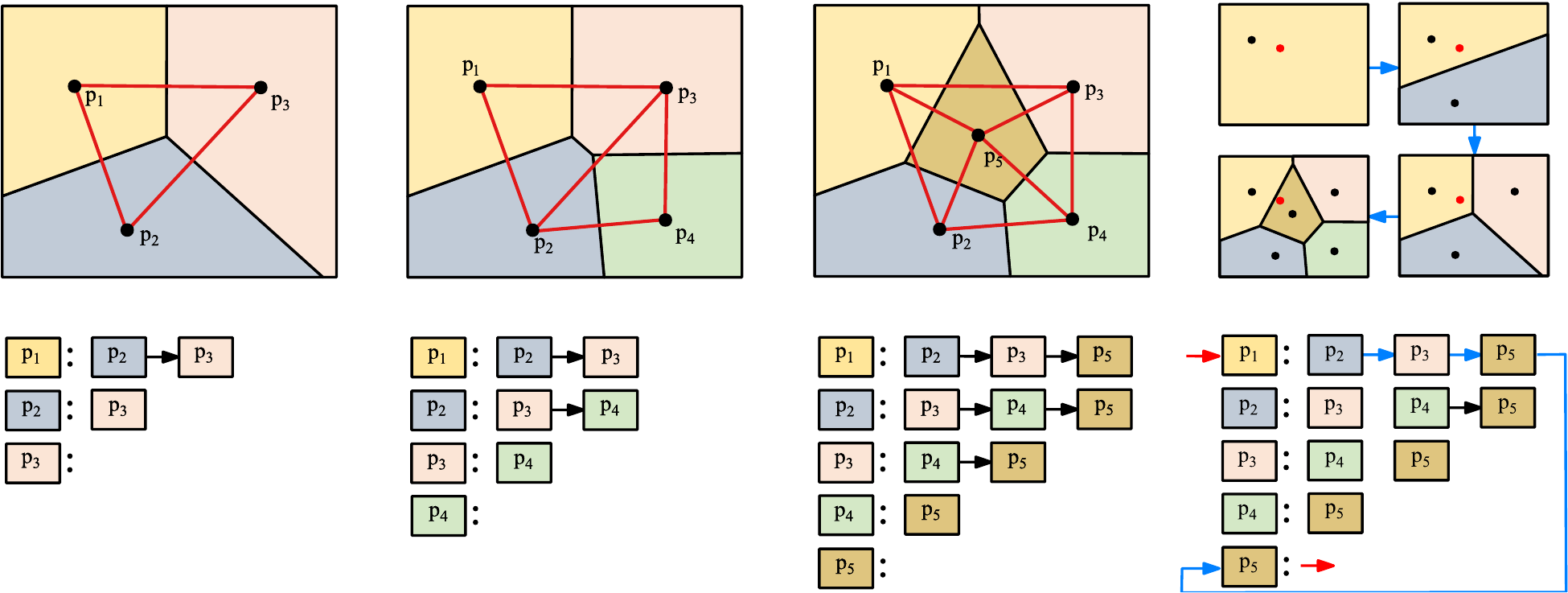}
 \put(10, -2){~\textbf{(a)}}
 \put(36, -2){~\textbf{(b)}}
 \put(61, -2){~\textbf{(c)}}
 \put(87, -2){~\textbf{(d)}}
\end{overpic}
\vspace{1mm}
\caption{
\CG{We consider an incremental construction process for the Voronoi diagram and its corresponding Delaunay triangulation of $\{p_i\}_{i=1}^n$. During this process, we append $p_j$ to $p_i$'s Query List if the insertion of $p_j$ modifies $p_i$'s Voronoi cell. In other words, within the Delaunay triangulation, $p_j$ and $p_i$ are adjacent neighbors. The rightmost figure illustrates this query process.}
}
\label{FIG:source}
	\vspace{-3mm}
\end{figure*}

 $\mathcal{L}$ is constructed simutaneously as the Voronoi diagram is incrementally built.
Initially, we have only one site $p_1$, leaving its Query List empty. Suppose that the addition of $p_{m+1}$ changes $p_i$'s cell in $\mathcal{V}_{m}$. We append $p_{m+1}$ to $p_i$'s Query List;
see Figure~\ref{FIG:source}. The Query Table $\mathcal{L}$ is completed once all the sites are inserted.
It is important to note that the order of elements stored in each $\mathcal{L}_i$ is consistent with the insertion order.

Since the Delaunay is the dual of the Voronoi diagram and encodes the adjacency relationships between Voronoi cells, to check
whether \( p_m \vdash \text{Cell}(p_i;\mathcal{V}_{m-1}) \),
this can be further converted into checking whether \( p_m \) and \( p_i \) are adjacent in the Delaunay constructed from \(\{p_i\}_{i=1}^m\), which is dual to $\mathcal{V}_{m}$.
\CG{Figure~\ref{FIG:source} illustrates the synchronized incremental construction process of both Voronoi diagrams and their dual Delaunay triangulations. As Delaunay triangulations explicitly encode structural relationships between points, obtaining neighboring points becomes straightforward, with most computational geometry libraries providing efficient interfaces for this purpose.}
Algorithm~\ref{alg:con} provides the pseudocode for the corresponding construction process.

\begin{algorithm}
\SetAlgoNoLine
\KwIn{Point cloud $\mathcal{P}=\{p_i\}_{i=1}^n$.}
\KwOut{Query Table $\{\mathcal{L}_i\}_{i=1}^n$.}
Initialize the Query Table $\mathcal{L}=\{\mathcal{L}_i\}_{i=1}^n$.

Insert \( p_1 \), initialize the Delaunay structure $\mathbf{D}$.

\ForEach{$i$ in $[2,n]$}{

Initialize the set $\omega$ as empty.

Insert \( p_i \) and update \( \mathbf{D} \).

Extract all points adjacent to \( p_i \), store them in $\omega$.

\ForEach{$p_x$ in $\omega$}{
Push $p_i$ to the back of $\mathcal{L}_x$.
}
}

\caption{Query Table Construction}
\label{alg:con}
\end{algorithm}

\subsection{Nearest Neighbor Query}
\label{sec:NearestNeighborQuery}
In the query phase, we initialize the nearest site as $p_1$, which implies $\Phi_{1}(q) = p_1$. Let $p_2$ be the first site in $p_1$'s Query List $\mathcal{L}_1$, and we visit $\mathcal{L}_1$ in order.
If $p_2$ provides a smaller distance than $p_1$, we switch to $p_2$'s Query List $\mathcal{L}_2$. Otherwise, we move to the next site in $\mathcal{L}_1$ to see if it can provide a smaller distance. See the rightmost figure of Figure~\ref{FIG:source} for how the query algorithm operates. 
See Algorithm~\ref{Algorithm:Query} for the pseudo-code of the query algorithm. 
\begin{algorithm}
\SetAlgoNoLine
\KwIn{Query Table $\{\mathcal{L}_i\}_{i=1}^n$ and query point $q$.}
\KwOut{The closest point to $q$.}
\SetKwFunction{Search}{getNearestNeighbor}
    \Search{$p_I=p_1$}{
    
      \ForEach{$p$ in $\mathcal{L}_I$}{
      \If{$\|q-p\| < \|q-p_I\|$}{
        \KwRet\ \Search{$p$}
      }
      }
      \KwRet\ $p_I$
    }
\caption{Nearest Neighbor Search}
\label{Algorithm:Query}
\end{algorithm}
\begin{figure}[h]
	\centering
\begin{overpic}
[width=.99\linewidth]{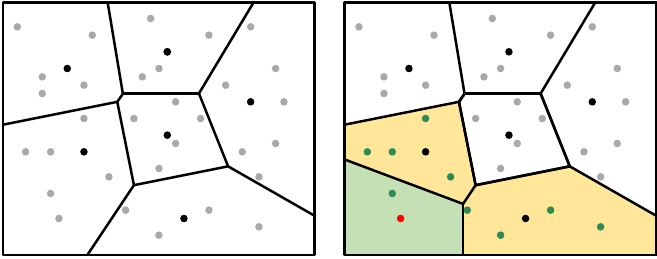}
\put(20,-4){~\textbf{(a)}}
\put(72,-4){~\textbf{(b)}}
\end{overpic}
\vspace{0.3mm}
\caption{
Coupling of Farthest Point sorting with incremental Voronoi Construction.
\CG{(a)}: Uninserted points (gray) are distributed within the Voronoi cells of the inserted points (black).
\CG{(b)}: When a new point is inserted, only the distance of the uninserted points within the green and yellow cells may need to be updated.
}
\label{Fig:FarestPointSampling}
\end{figure}

\subsection{Insertion Reordering Based on Farthest-Point Sorting}
\label{sec:FPS}
Sections~\ref{sec:queryTableConstruction} and ~\ref{sec:NearestNeighborQuery} describe the algorithm for constructing the Query Table and performing nearest neighbor searches. The algorithm assumes a specific order of the point cloud and incrementally constructs the Voronoi / Delaunay based on this order. 
\CG{When the point cloud is unordered, we can sort the points in a specific manner to enhance the efficiency of either the preprocessing or query phase.}
\CG{Intuitively, the more uniformly the inserted point cloud is distributed in space, the faster the query speed, as it facilitates a transition from large to small spatial jumps during the query process.}

\CG{Farthest point sampling (FPS) is a technique that iteratively selects points from a dataset such that each newly selected point is maximally distant from the previously chosen points. This process can be formalized as: \begin{equation} p_{i+1} = \arg\max_{p \in P} \left( \min_{p_j \in S} d(p, p_j) \right), \end{equation} where $P$ represents the point cloud, $S$ denotes the already selected points, $d(p, p_j)$ is the distance between points, and $p_{i+1}$ is the next selected point that maximizes its minimum distance to any point in $S$. This strategy ensures well-distributed point selection, making it effective for shape approximation and surface representation.}

Drawing inspiration from this, we set the number of sampled points equal to the total number of points in the cloud and use the sampling order to incrementally construct the \CG{Delaunay triangulation}. Notably, by integrating this approach into our construction process, we developed a farthest point sampling strategy with reduced complexity, leading to faster sampling times.

An important step in the farthest point sampling process involves calculating the distance between the newly sampled point and all remaining points, followed by updating the distance information. However, in most cases, the distances for the majority of the remaining points do not require updating, leading to a significant waste of computational resources. By integrating this process with our construction method, we can optimize the calculations by updating the distances for only a small subset of remaining points that are likely to change. This approach improves the algorithm's efficiency. 

As an example, consider a state from the incremental construction of the Voronoi diagram, as shown in Figure~\ref{Fig:FarestPointSampling}(a). The black points represent the points that have already been inserted, while the gray points denote the remaining points waiting to be inserted. When a new point is inserted based on the farthest point sampling strategy, as illustrated in Figure~\ref{Fig:FarestPointSampling}(b), the red point represents the newly sampled point from the gray points, and the green area corresponds to its Voronoi cell.
A key observation is that only the distances of the remaining points located within the cells adjacent to the green cell (specifically, those within these cells prior to the insertion of the red point) need to be updated. Since the white Voronoi cell does not change during the insertion of new points, this implies that the nearest neighbors of the uninserted points within it remain unchanged. 
This significantly reduces the computational cost of sampling.

\CG{
The remaining task involves two parts: dynamically tracking the remaining points contained within the cells of the already inserted points during the incremental construction of the Voronoi diagram, and dynamically obtaining the next sampling point.
For the first task, we dynamically maintain a vector for each inserted point, storing the remaining points that are nearest to that point. We update these vectors dynamically during the incremental Delaunay construction process. For the second task, a segment tree~\cite{mir_mods_00001178} provides an excellent data structure that supports both single-point value updates within the interval and maximum value queries within same interval, both in logarithmic time. Therefore, during incremental Delaunay construction, we obtain all remaining points requiring distance updates through vector updates, update them in the segment tree (with inserted points set to negative infinity), and then leverage the segment tree's logarithmic-time maximum query capability to retrieve the next farthest sampling point.
}

\subsection{Two Delaunay Construction Strategies}
\CG{
For the construction of Delaunay triangulations, numerous algorithms exist; we employ the Bowyer–Watson algorithm~\cite{10.1093/comjnl/24.2.162}, which performs incremental Delaunay construction through sequential point insertion. When inserting a new point, the algorithm first locates the Delaunay tetrahedron containing the point's spatial location, then performs edge flipping operations based on the empty sphere property until the Delaunay criteria are satisfied. 
In our framework, different insertion orders for the Delaunay construction can achieve either faster construction speed or improved query performance. We introduce two strategies adopted in this work:
}

\textbf{Prioritizing Construction Speed.}
\CG{
One time-consuming operation during Delaunay construction is locating the tetrahedron containing the newly inserted point. To accelerate Delaunay construction, a common approach involves pre-sorting the input point cloud using space-filling curves before performing incremental construction. This strategy ensures that the point insertion order exhibits spatial adjacency properties, enabling newly inserted points to quickly propagate from the previous insertion location to locate the tetrahedron containing the new point. Additionally, this approach provides excellent spatial locality during the construction process.

We adopt the point cloud sorting strategy from CGAL~\cite{cgal:dd-ss-24b}, specifically the \texttt{spatial\_sort} function, which first organizes the point set into random buckets of increasing sizes, with Hilbert sorting applied only within each bucket. Our experiments demonstrate that this strategy achieves fast Delaunay construction speed with minimal impact on query performance.
}

\textbf{Prioritizing Query Performance.}
\CG{
Farthest Point Sorting ensures that inserted points are distributed as uniformly as possible throughout the full point cloud, which intuitively leads to faster query speeds—a hypothesis validated by our experiments. The farthest point sampling strategy has been detailed in Section~\ref{sec:FPS}. Here, we primarily focus on how our algorithm can be integrated to accelerate the Delaunay construction process under the farthest point sampling ordering.

Our algorithm possesses an important property: as illustrated in the Figure~\ref{FIG:source} (b,c), before inserting the 5th point, the query table for the first four points has already been constructed (shown in the bottom portion of the Figure~\ref{FIG:source} (b) ), enabling queries for the nearest point among the first 4 points. Consequently, when inserting point $p_k$, we can already utilize Algorithm~\ref{Algorithm:Query} to query the nearest point to $p_k$ within $\{p_i\}_{i=1}^{k-1}$.

As previously mentioned, one time-consuming operation during Delaunay construction is locating the tetrahedron containing the newly inserted point. Therefore, when inserting the $k$-th point, we can quickly determine the nearest point to $p_k$ from $\{p_i\}_{i=1}^{k-1}$. This nearest point then serves as the starting location for efficiently locating the tetrahedron containing $p_k$, significantly accelerating the Delaunay construction process.
}



\subsection{Time Complexity Analysis}

\textbf{Preprocessing.}
\CG{Our preprocessing phase involves the incremental construction of the Delaunay triangulation, during which we store the requisite information for subsequent queries. Consequently, the preprocessing time complexity aligns with that of Delaunay triangulation construction.

On average, the time complexity for constructing a Delaunay triangulation in 3D is $O(n \log n)$~\cite{10.5555/261226}, while the worst-case time complexity is $O(n^2)$~\cite{10.1145/777792.777823}. However, this worst-case scenario is rarely encountered in typical construction processes.}


\textbf{Querying.}
\CG{Due to the difficulty in evaluating the impact of point cloud distribution on Delaunay construction, providing a rigorous theoretical proof for average-case complexity remains challenging.
Therefore, we present an experimental average complexity based on empirical observations, followed by a theoretical worst-case complexity analysis.
The experimental section validates the correctness of our experimental average complexity analysis.}

\paragraph{Experimental Average Complexity}
\CG{When querying the nearest neighbor for point $q$, the complexity of the query phase depends on the number of times the Voronoi cell containing $q$ undergoes shrinkage during the incremental construction process. Each time the Voronoi cell containing $q$ shrinks due to the insertion of a new point, a comparison operation is required to potentially update the nearest neighbor.

More specifically, assume $\Phi_m(q)$ represents the current nearest neighbor of $q$ among the first $m$ points, and a new point $p_{m+1}$ is inserted. 
Based on empirical observations, on average, the Voronoi cell of $p_{m+1}$ is adjacent to a constant number of existing cells, denoted as $c$~\cite{mei2018degreedistributiondelaunaytriangulations}.
Consequently, the probability that $p_{m+1}$ is adjacent to $\Phi_m(q)$ in the Delaunay triangulation is approximately $c/m$, which corresponds to the likelihood of requiring a comparison operation.

Thus, the average time complexity of the nearest neighbor query phase is: \begin{equation} 
O\left(\sum_{i=1}^{n-1} \frac{c}{i}\right) = c \times O(\log n) = O(\log n). 
\end{equation}
}
\CG{
\paragraph{Theoretical Worst-Case Complexity}
\CG{The theoretical worst-case complexity of our algorithm is $O(n)$, as it relies on pairwise point comparisons and eliminates at least one invalid candidate with each comparison. 
A typical example is when all points in 3D space lie on a straight line, ordered sequentially as $p_1, p_2, \ldots, p_n$, with Delaunay construction following this insertion order. In this degenerate configuration, each point $p_k$ encroaches only upon $p_{k-1}$, creating a chain-like structure that forces linear traversal during nearest neighbor queries, thereby degrading the algorithm to $O(n)$ complexity.}
}
\begin{figure}[h]
	\centering
\begin{overpic}
[width=.99\linewidth]{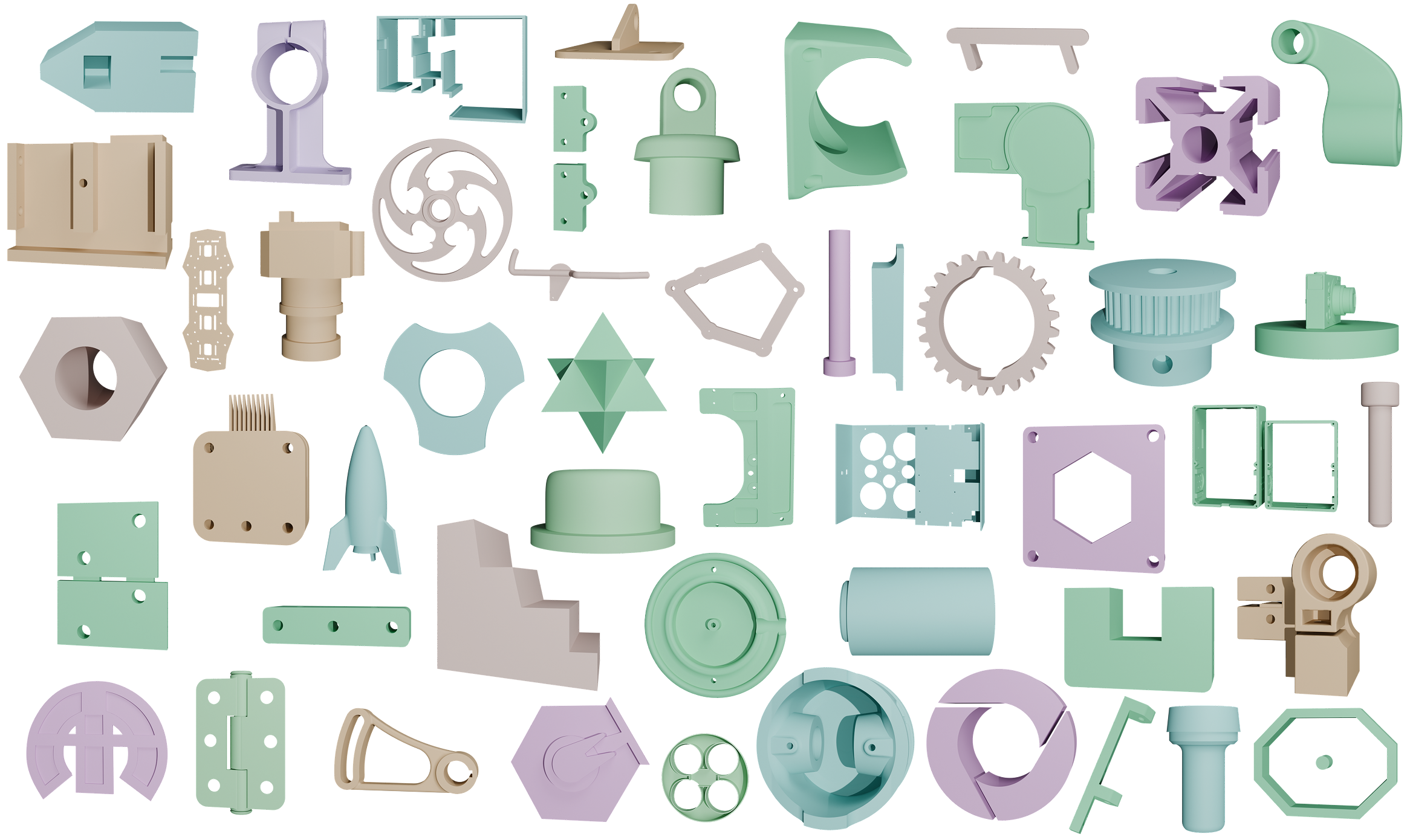}
\end{overpic}
\vspace{-8mm}
\caption{
\CG{Representative 3D models from the ABC dataset~\cite{Koch_2019_CVPR}.}
}
\label{Fig:abc_gallery}
\end{figure}

\begin{figure}[h]
	\centering
\begin{overpic}
[width=.99\linewidth]{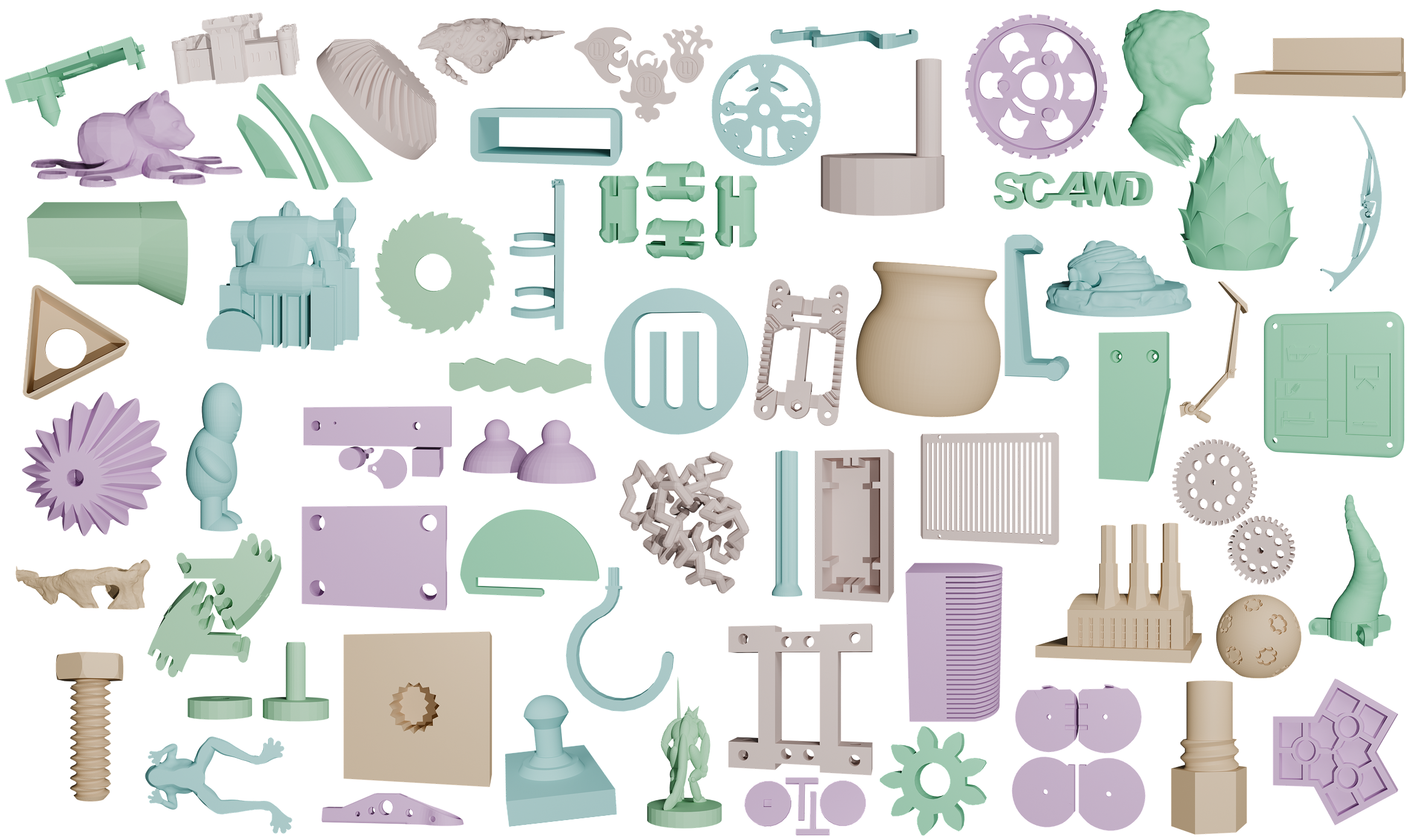}
\end{overpic}
\vspace{-8mm}
\caption{
\CG{Representative 3D models from the Thingi10K dataset~\cite{Thingi10K}.}
}
\label{Fig:thingi10k_gallery}
\end{figure}

\begin{figure}[h]
	\centering
\begin{overpic}
[width=.99\linewidth]{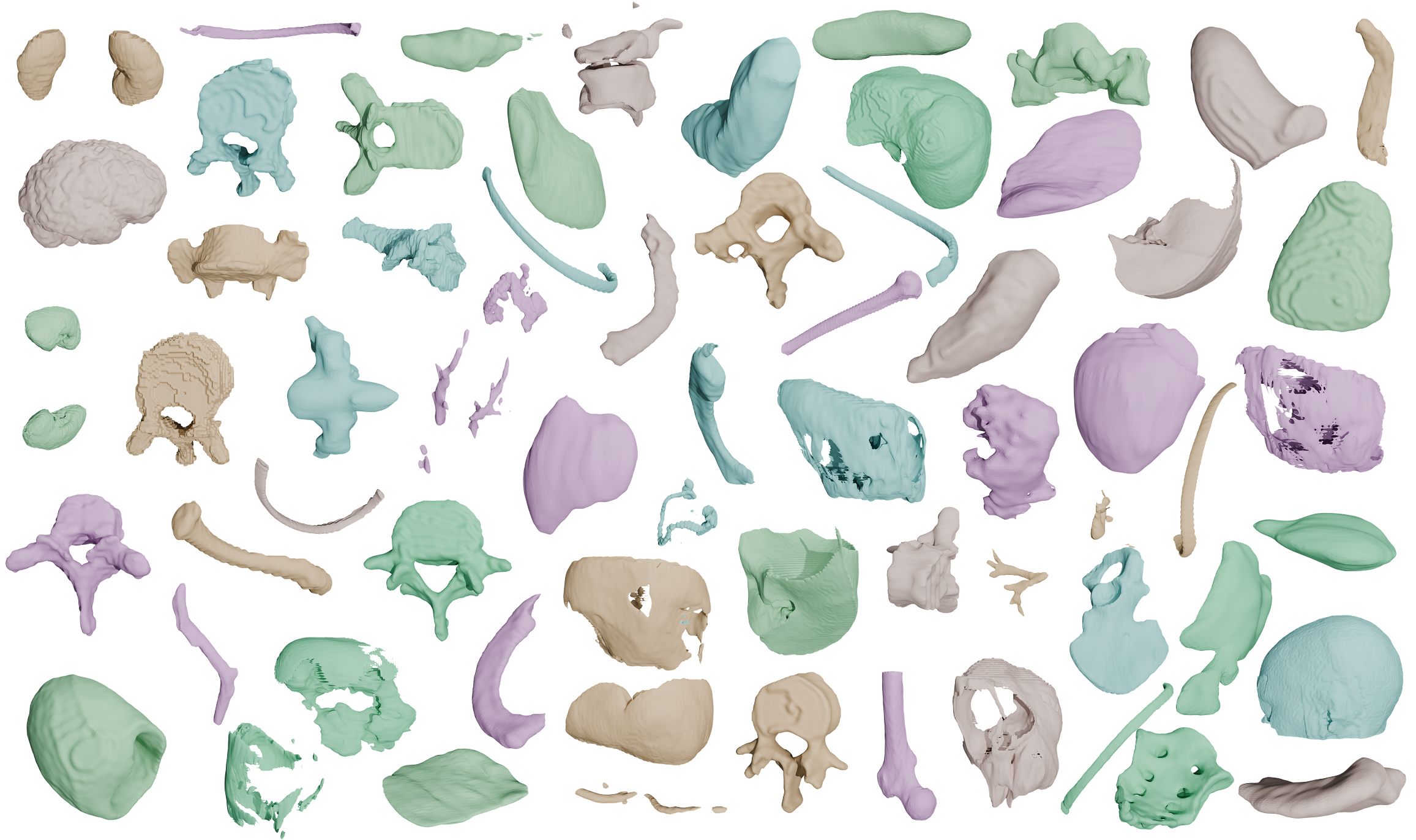}
\end{overpic}
\vspace{-8mm}
\caption{
\CG{Representative 3D models from the MedShapeNet dataset~\cite{li2023medshapenet}.}
}
\label{Fig:medshapeNet_gallery}
\end{figure}

\section{Results}
\textbf{Platform.}
\CG{Our algorithm was implemented in C++ and evaluated on a Mac mini equipped with an M4 CPU, 16GB of RAM, running the macOS operating system. We utilized CGAL~\cite{cgal:hs-chdt3-24a} for Delaunay triangulation computations and  employed CGAL's multi-scale sorting approach and farthest point sorting as the insertion order strategies.}

\textbf{Relevant approaches for comparison.}
\CG{We selected six classic methods as comparison baselines, and together with our proposed method (incorporating two different pre-sorting strategies), we evaluated a total of eight approaches.}
\CG{
\begin{enumerate}
    \item KD Tree: From the nanoflann library~\cite{blanco2014nanoflann}, with the maximum number of leaf nodes set to the default value of 10, which achieves nearly optimal query performance according to the benchmark results provided by the official repository.
    \item R* Tree: From Boost library~\cite{BoostLibrary}, with the maximum number of elements per node set to 10, consistent with the KD Tree configuration.
    \item Octree: From PCL library~\cite{Rusu_ICRA2011_PCL}, with the resolution parameter set to 3 times the inter-point distance.
    \item BD Tree: From ANN library, with the maximum number of elements in leaf node set to 10, and we only use it for exact nearest neighbor queries.
    \item Voronoi: From CGAL library, utilizing its provided \texttt{nearest\_vertex} function for nearest neighbor queries.
    \item Grid: Our own implementation, with cell side length set to 3 times the inter-point distance.
    \item Ours$^1$: Our method, using CGAL's multi-scale sorting method \texttt{spatial\_sort} for incremental Delaunay construction.
    \item Ours$^2$: Our method, using farthest point sorting strategy for incremental Delaunay construction.
\end{enumerate}
}
\CG{Additionally, it is worth noting that the Grid method exhibits slower performance when there is a considerable distance between query points and target points. The Voronoi-based method is also relatively slow. Therefore, we only include results from these two methods in certain experiments.}

\textbf{Datasets and Query Point Generation.}
\CG{We selected three datasets and several standard models for algorithm evaluation. 
For nearest neighbor queries, we utilize the vertices of these 3D models as the point cloud data.
After filtering out models that are excessively large or small, the datasets cover the following ranges:
\begin{enumerate}
    \item ABC Dataset~\cite{Koch_2019_CVPR}: We tested on a subset of 10,000 models, with the final dataset covering 600 to 600,000 points. Figure~\ref{Fig:abc_gallery} shows representative models from this dataset.
    \item Thingi10K Dataset~\cite{Thingi10K}: Originally containing 10,000 models, the filtered dataset covers 400 to 140,000 points. Figure~\ref{Fig:thingi10k_gallery} displays sample models from this collection.
    \item MedShapeNet Dataset~\cite{li2023medshapenet}: We evaluated on a subset of 10,000 models, with the processed dataset covering 200 to 150,000 points. Figure~\ref{Fig:medshapeNet_gallery} illustrates representative models from this dataset.
\end{enumerate}

For query point generation, we randomly distribute 1 million query points within the 2× axis-aligned bounding box for each test case and perform nearest neighbor queries.
}

\begin{figure}[t]
	\centering
\begin{overpic}
[width=.99\linewidth]{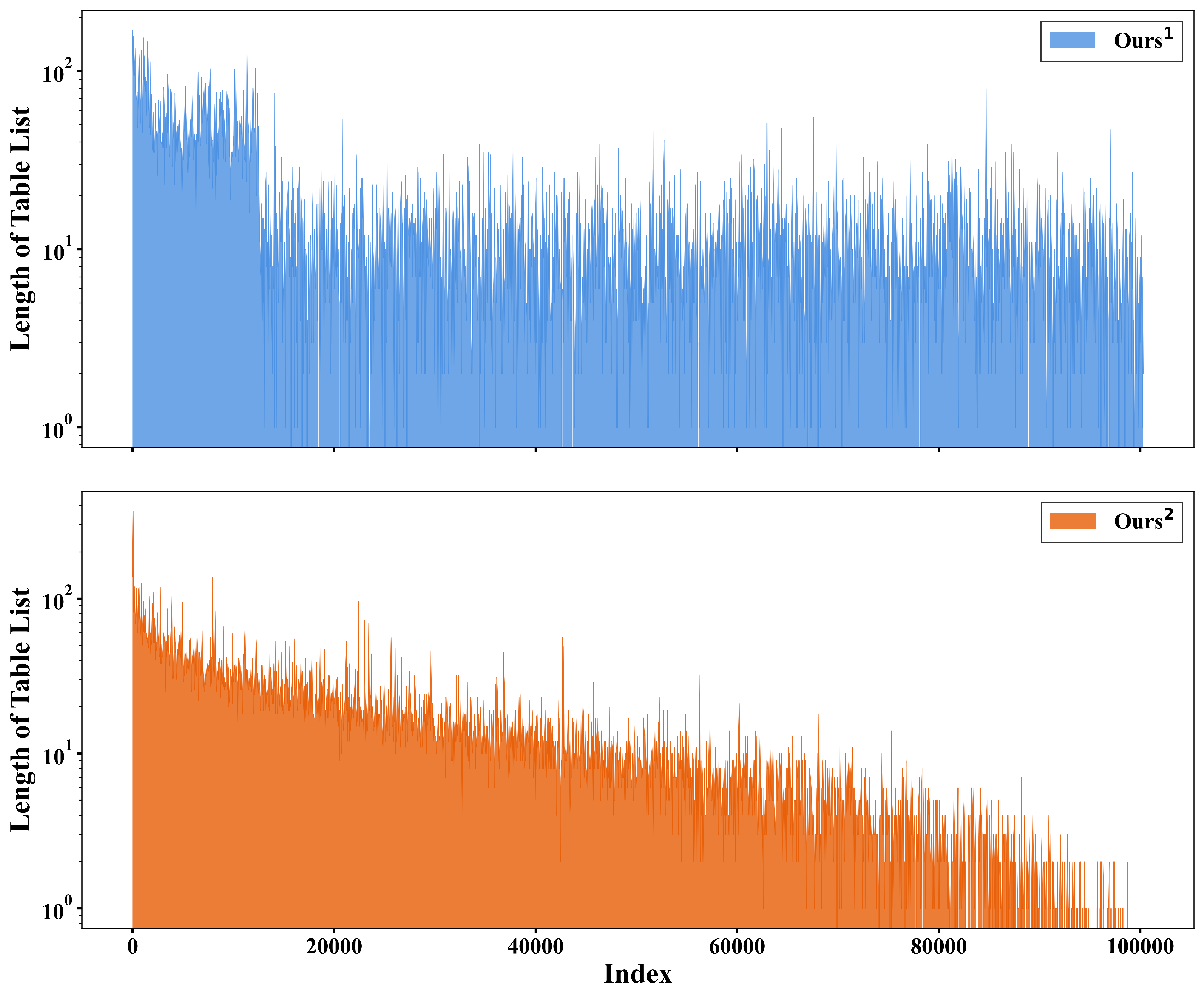}
\end{overpic}
\vspace{-8mm}
\caption{
The length of the Query List for different index points. Top: \CG{CGAL's multi-scale sorting}. Bottom: farthest point sorting.
}
\label{Fig:LengthofTableList}
\end{figure}

\begin{figure}[h]
	\centering
\begin{overpic}
[width=.99\linewidth]{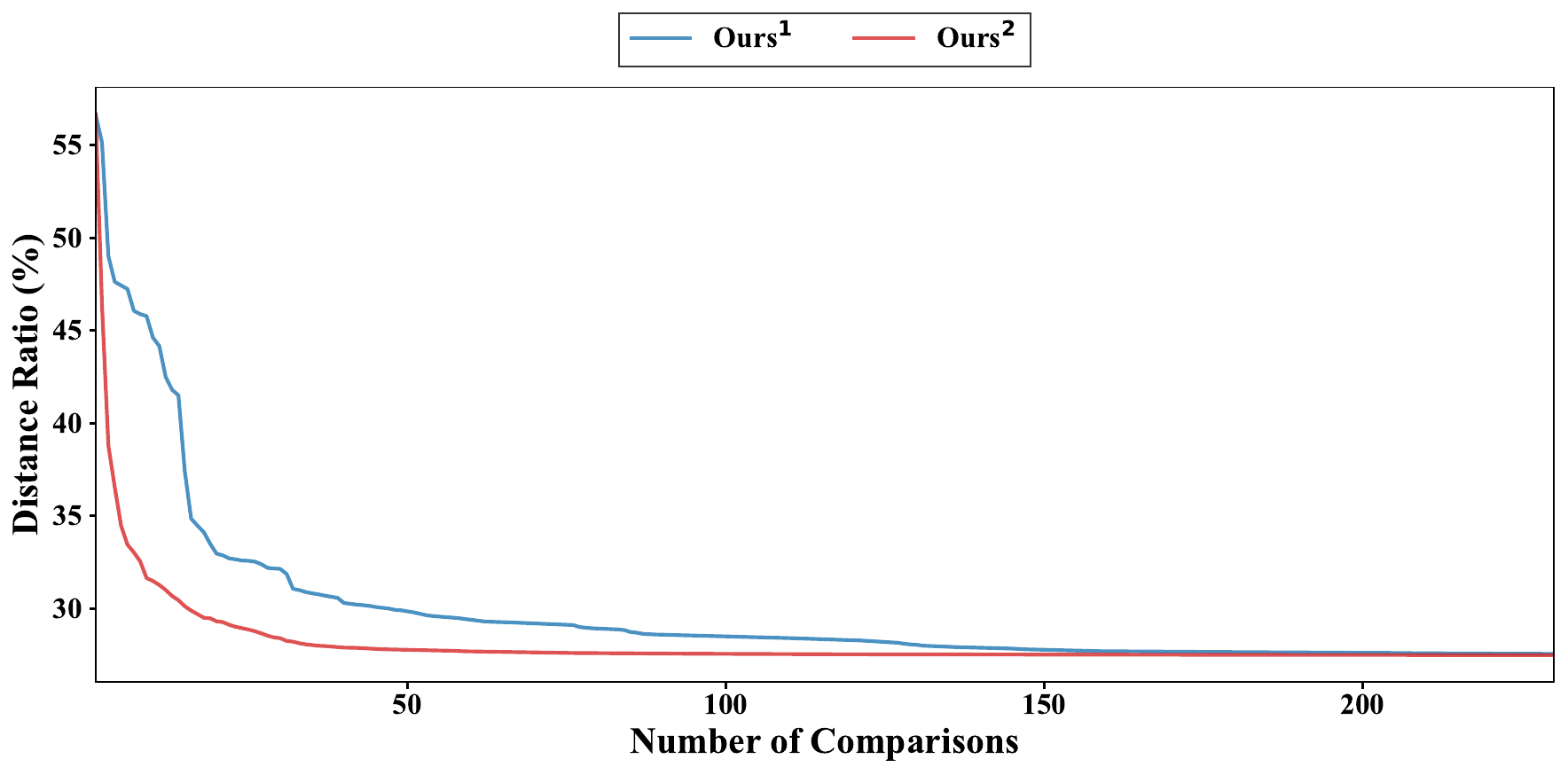}
\end{overpic}
\vspace{-8mm}
\caption{
\CG{Query process convergence under two insertion strategies. Average nearest distance (normalized by point cloud diagonal) versus number of comparisons, averaged over one million queries.}
}
\label{Fig:QueryProcessConvergence}
\end{figure}

\begin{table}[]
\begin{center}

\caption{\CG{The average (Avg) and variance (Var) of the length of the query list and the number of comparisons required per query on classical models.}}
\label{table:TestPoints}
\vspace{-2mm}
\setlength{\tabcolsep}{0.1cm} 
\renewcommand{\arraystretch}{1.3} 
\resizebox{1\linewidth}{!}
{
\begin{tabular}{ccccccccl}
\toprule
\multicolumn{1}{l}{}                                  &                             & Camel  & Bunny  & Dragon & Kitten & Armadillo & Lucy    & Sponza  \\ \midrule
\multicolumn{2}{c|}{Vertices}                                                       & 28934  & 72911  & 100313 & 291023 & 726367    & 1018219 & 1313504 \\ \midrule
\multicolumn{1}{c|}{\multirow{3}{*}{$\text{Ours}^1$}} & \multicolumn{1}{c|}{Avg}    & 16.753 & 16.445 & 16.371 & 15.508 & 15.864    & 16.106  & 15.765  \\
\multicolumn{1}{c|}{}                                 & \multicolumn{1}{c|}{Var}    & 20.099 & 19.281 & 16.619 & 19.181 & 19.026    & 19.301  & 18.713  \\
\multicolumn{1}{c|}{}                                 & \multicolumn{1}{c|}{Tested} & 188.0  & 183.96 & 240.15 & 214.39 & 274.27    & 304.22  & 339.05  \\ \hline
\multicolumn{1}{c|}{\multirow{3}{*}{$\text{Ours}^2$}} & \multicolumn{1}{c|}{Avg}    & 14.414 & 14.340 & 14.197 & 14.115 & 14.246    & 14.333  & 13.638  \\
\multicolumn{1}{c|}{}                                 & \multicolumn{1}{c|}{Var}    & 16.783 & 15.794 & 16.355 & 16.207 & 15.947    & 16.016  & 15.927  \\
\multicolumn{1}{c|}{}                                 & \multicolumn{1}{c|}{Tested} & 144.87 & 136.28 & 162.68 & 152.35 & 193.15    & 211.39  & 235.85  \\ \bottomrule
\end{tabular}
}
\end{center}
\end{table}
\begin{figure*}[h]
	\hspace{-3mm} 
\begin{overpic}
[width=0.99\linewidth]{imgs/processing_macos_dataset.pdf}
\end{overpic}
\vspace{-3mm}
\caption{\CG{
Preprocessing time comparison across ABC, Thingi10K, and MedShapeNet datasets (from left to right).}}
\label{Fig:DataProcessingTimeOnThreeDatasets}
\end{figure*}

\begin{figure}[h]
	\hspace{-3mm} 
\begin{overpic}
[width=0.99\linewidth]{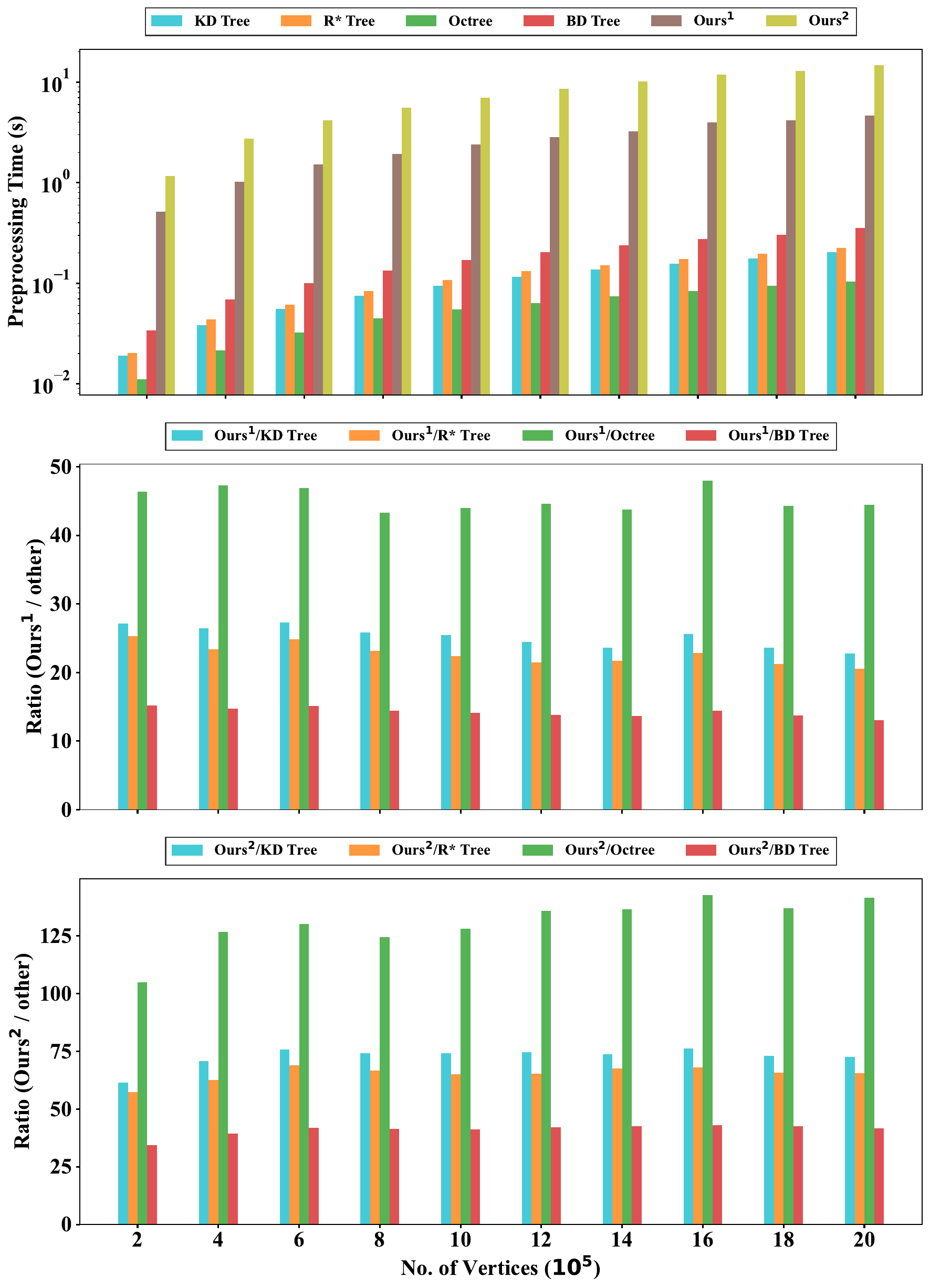}
\end{overpic}
\vspace{-3mm}
\caption{
\CG{Preprocessing cost comparison of various methods on the Dragon model across different resolutions. Top: absolute preprocessing time. Middle and Bottom: relative preprocessing cost ratios between methods.}
}
\label{Fig:DragonVaryVerticesProcessing}
\end{figure}

\subsection{Query Table}
\textbf{Analysis of Query List Length.}
\CG{Recall that each vertex maintains a Query List, which is a vector storing the subsequently inserted points that become adjacent to the current vertex in the Delaunay triangulation during the incremental construction process.
It is evident that points inserted earlier experience more subsequent insertions, resulting in longer Query Lists as they accumulate more neighbors over time.}
Therefore, to investigate how the length of the Query List varies with different insertion indexes, we conducted a study using the $100$K-vertex dragon point cloud as an example. 
As shown in Figure~\ref{Fig:LengthofTableList}, we have plotted the length of the query list for each point with respect to its insertion order.
\CG{The results reveal that two different insertion orders lead to distinctly different query length distribution trends. CGAL's multi-scale sorting method exhibits a clear two-stage characteristic: the first stage demonstrates significantly longer average query lengths compared to the second stage, while the query lengths fluctuate around their respective average values within each stage.
In contrast, the query lengths obtained through farthest point sampling show a steady declining trend.}

\textbf{Number of Comparisons.}
As shown in Algorithm~\ref{Algorithm:Query}, the nearest neighbor search problem is transformed into the task of calculating distances between points.
The number of distance comparisons between points during the query process partially reflects the efficiency of the nearest neighbor search.
We tested several classical models and recorded the number of comparisons, along with the mean and variance of the Query Lists, as shown in Table~\ref{table:TestPoints}.
It can be observed that, after utilizing the farthest point sorting method, both the average length (Avg) and variance (Var) of the Query Lists decreased, accompanied by a reduction in the average number of comparisons during the nearest neighbor search process.

\textbf{Query Process Convergence.}
\CG{
In accordance with Algorithm~\ref{Algorithm:Query}, our method iteratively evaluates target points during query execution while dynamically updating the current nearest distance estimate.
To quantitatively assess the influence of varying insertion orders on query efficiency, we conducted a comprehensive evaluation using the 100K dragon model. Specifically, we generated 1 million query points within the search space and executed nearest neighbor queries for each point.
During execution, we systematically recorded the current nearest distance at each comparison step, subsequently computing the average distance across all queries per step. Figure~\ref{Fig:QueryProcessConvergence} demonstrates the convergence behavior of this average nearest distance (normalized as a percentage of the point cloud's diagonal length) as a function of comparison count under two distinct insertion ordering strategies.
The results reveal that the Farthest Point Sampling strategy facilitates substantially accelerated and more consistent convergence throughout the query process, empirically validating its efficacy in enhancing search performance.
}
\begin{table}[]
\begin{center}
\caption{\CG{Preprocessing time~($s$) on classical models}}
\label{table:prcessingTimeOnModels}
\vspace{-2mm}
\setlength{\tabcolsep}{0.1cm} 
\renewcommand{\arraystretch}{1.3} 
\resizebox{1\linewidth}{!}
{
\begin{tabular}{c|ccccccl}
\toprule
           & Camel & Bunny & Dragon & Kitten & Armadillo & Lucy    & Sponza  \\ \midrule
Vertices   & 28934 & 72911 & 100313 & 291023 & 726367    & 1018219 & 1313504 \\ \midrule
KD-tree    & 0.003 & 0.006 & 0.009  & 0.025  & 0.068     & 0.098   & 0.123   \\
R$^*$-tree & 0.004 & 0.007 & 0.010  & 0.028  & 0.078     & 0.109   & 0.170   \\
Octree     & 0.002 & 0.003 & 0.004  & 0.011  & 0.033     & 0.047   & 0.055   \\
BD-tree    & 0.005 & 0.010 & 0.014  & 0.044  & 0.118     & 0.163   & 0.198   \\
Grid       & 0.002 & 0.004 & 0.008  & 0.024  & 0.313     & 0.362   & 0.053   \\
Voronoi    & 0.057 & 0.143 & 0.189  & 0.517  & 1.359     & 1.943   & 2.466   \\
Ours$^1$   & 0.072 & 0.182 & 0.249  & 0.685  & 1.736     & 2.481   & 3.170   \\
Ours$^2$   & 0.103 & 0.327 & 0.466  & 1.698  & 5.005     & 7.454   & 9.602   \\ \bottomrule
\end{tabular}
}
\end{center}
\end{table}

\subsection{Preprocessing Cost}
The preprocessing time of our algorithm is primarily concentrated in the Delaunay construction phase, as it involves significant computational overhead in building and managing the complex graph structure. In this section, we present the preprocessing times of different algorithms across several classic models in Table~\ref{table:prcessingTimeOnModels}. \CG{Figure~\ref{Fig:DragonVaryVerticesProcessing} further illustrates how preprocessing time scales with point cloud resolution using the Dragon model as a representative example.
We extend this evaluation to larger datasets in Figure~\ref{Fig:DataProcessingTimeOnThreeDatasets}, which demonstrates preprocessing performance across the ABC~\cite{Koch_2019_CVPR}, Thingi10k~\cite{Thingi10K}, and MedShapeNet~\cite{li2023medshapenet} datasets.
}
It can be observed that at this stage, our method currently exhibits longer preprocessing times compared to existing nearest neighbor search algorithms, particularly at higher resolutions, indicating areas that require further optimization.

\subsection{Query Performance}
\begin{figure}[h]
	\hspace{-3mm} 
\begin{overpic}
[width=0.99\linewidth]{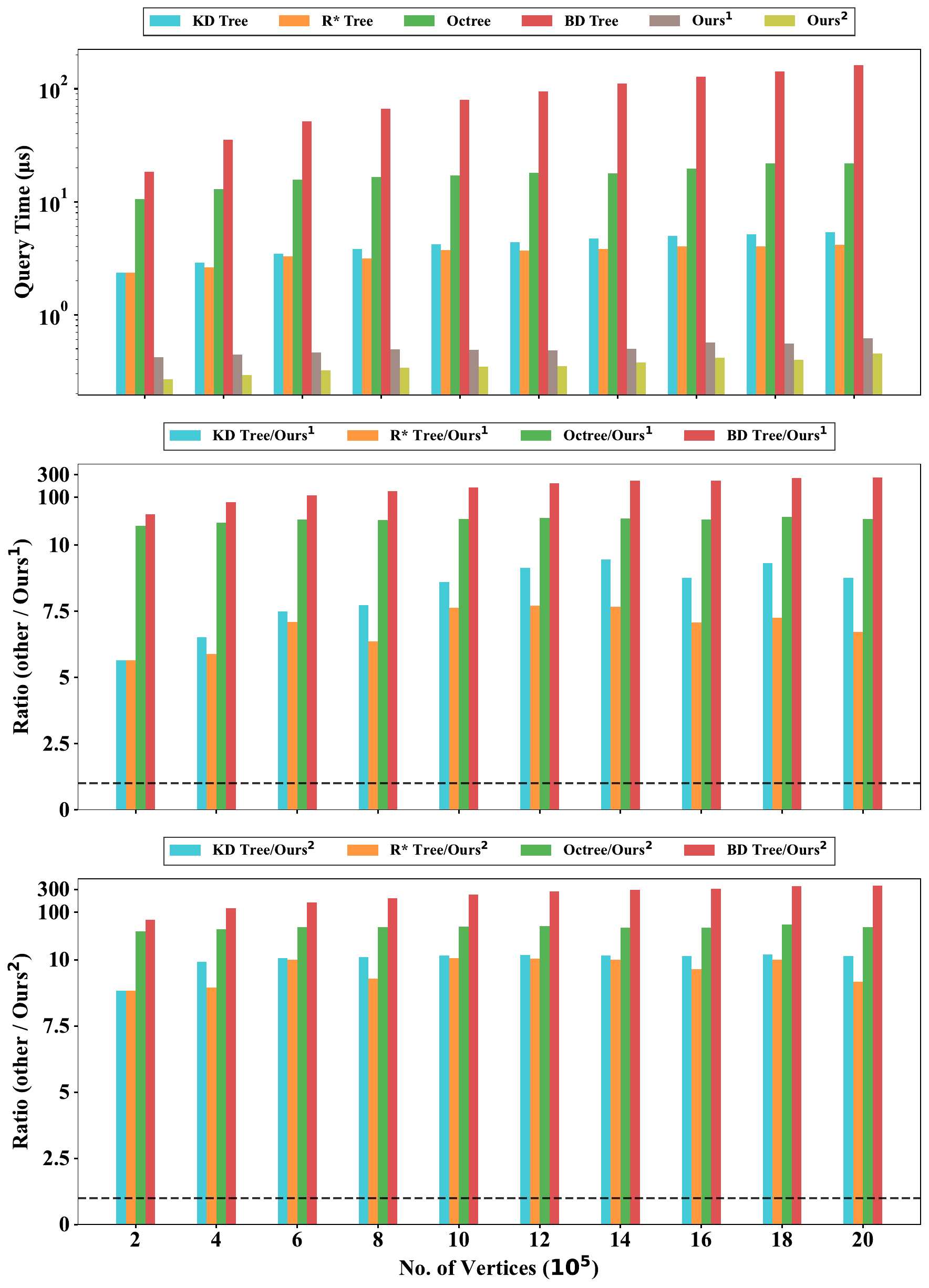}
\end{overpic}
\vspace{-3mm}
\caption{
\CG{Average query time on the Dragon model across varying resolutions. Top: Absolute query time ($\mu$s) for different methods. Middle and Bottom: Performance ratios between methods, with dashed line indicating equal performance (ratio = 1).}
}
\label{Fig:DragonVaryVerticesQuery}
\end{figure}

\CG{
Query efficiency constitutes the primary advantage of our proposed approach. To systematically evaluate this performance benefit, we conducted a multi-faceted analysis across various datasets and resolutions.

Using the Dragon dataset as a controlled benchmark, we measured query costs with respect to point cloud resolution, as illustrated in Figure~\ref{Fig:DragonVaryVerticesQuery}. This analysis demonstrates a consistent and significant performance advantage for our method across all tested resolutions. Table~\ref{table:QueryTimeOnModels} complements this analysis by providing detailed query performance comparisons across a diverse set of classical models at their standard resolutions, further validating our method's superior efficiency.

To establish the generalizability of these performance advantages, we expanded our evaluation to large-scale datasets with diverse geometric characteristics: Thingi10K~\cite{Thingi10K}, ABC~\cite{Koch_2019_CVPR}, and MedShapeNet~\cite{li2023medshapenet}. Figure~\ref{Fig:DatasetQueryTime} summarizes these comprehensive results, revealing consistent performance patterns across all tested repositories. Notably, our method achieves query acceleration factors of 1-10× compared to current state-of-the-art approaches, with substantial improvements observed across various model types.}

\subsection{Memory Usage}
\CG{Space complexity represents another critical dimension for algorithm evaluation. To provide a comprehensive assessment, we analyzed the memory requirements of our approach alongside existing methods. The primary memory consumption in our algorithm stems from three components: incremental Delaunay triangulation construction, Query Table storage, and segment tree structures when implementing farthest point sampling.

It should be noted that precise memory quantification presents inherent challenges due to system-level memory management mechanisms, resulting in measurement fluctuations. For statistical robustness, our analysis excludes data points with memory utilization below 0.2M.
Figure~\ref{Fig:datasetMemoryUsed} presents the memory consumption of our algorithm across the ABC, Thingi10K, and MedShapeNet datasets.

The empirical results demonstrate that while our method exhibits moderately increased memory requirements compared to traditional approaches, this additional memory overhead remains within acceptable bounds given the significant query performance advantages. This trade-off between space complexity and query efficiency represents a favorable compromise for applications where query speed is paramount.
}

\begin{figure*}[h]
	\hspace{-3mm} 
\begin{overpic}
[width=0.99\linewidth]{imgs/query_macos_dataset.pdf}
\end{overpic}
\vspace{-3mm}
\caption{\CG{
Query time comparison across ABC, Thingi10K, and MedShapeNet datasets (from left to right). Dashed line indicates equal performance (ratio = 1).}}
\label{Fig:DatasetQueryTime}
\end{figure*}

\begin{figure*}[h]
	\hspace{-3mm} 
\begin{overpic}
[width=0.99\linewidth]{imgs/memory_usage.pdf}
\end{overpic}
\vspace{-3mm}
\caption{\CG{Memory utilization comparison across ABC, Thingi10K, and MedShapeNet datasets.}}
\label{Fig:datasetMemoryUsed}
\end{figure*}

\begin{table}[]
\caption{\CG{Average query time ($\mu$s) comparison across classical models.
The \textbf{highest} and \underline{second}-fastest query times are highlighted in bold and underlined, respectively}}
\label{table:QueryTimeOnModels}
\vspace{-2mm}
\setlength{\tabcolsep}{0.1cm} 
\renewcommand{\arraystretch}{1.3} 
\resizebox{1\linewidth}{!}
{
\begin{tabular}{c|ccccccc}
\hline
         & Camel          & Bunny          & Dragon         & Kitten             & Armadillo          & Lucy               & Sponza             \\ \hline
Vertices & 28934          & 72911          & 100313         & 291023             & 726367             & 1018219            & 1313504            \\ \hline
KD-tree  & 1.109          & 2.542          & 1.585          & 5.402              & 6.003              & 4.296              & 1.815              \\
R*-tree  & 1.450          & 2.584          & 1.808          & 4.974              & 4.967              & 3.616              & 1.446              \\
Octree   & 5.060          & 8.475          & 5.719          & 17.06              & 17.21              & 14.89              & 10.22              \\
BD-tree  & 2.344          & 4.721          & 5.715          & 13.11              & 78.04              & 23.02              & 2.674              \\
Grid     & 52.81          & 352.5          & 867.1          &  $>2k$ &  $>2k$ &  $>2k$ &  $>2k$ \\
Voronoi  & 6.815          & 9.965          & 5.692          & 27.01              & 26.48              & 39.10              & 36.92              \\
Ours$^1$ & {\ul 0.285}    & {\ul 0.322}    & {\ul 0.343}    & {\ul 0.416}        & {\ul 0.465}        & {\ul 0.501}        & {\ul 0.716}        \\
Ours$^2$ & \textbf{0.201} & \textbf{0.235} & \textbf{0.213} & \textbf{0.322}     & \textbf{0.377}     & \textbf{0.401}     & \textbf{0.617}     \\ \hline
\end{tabular}
}
\end{table}

\subsection{Performance on Uniform Distributions}

\begin{figure}[h]
	\hspace{-3mm} 
\begin{overpic}
[width=0.99\linewidth]{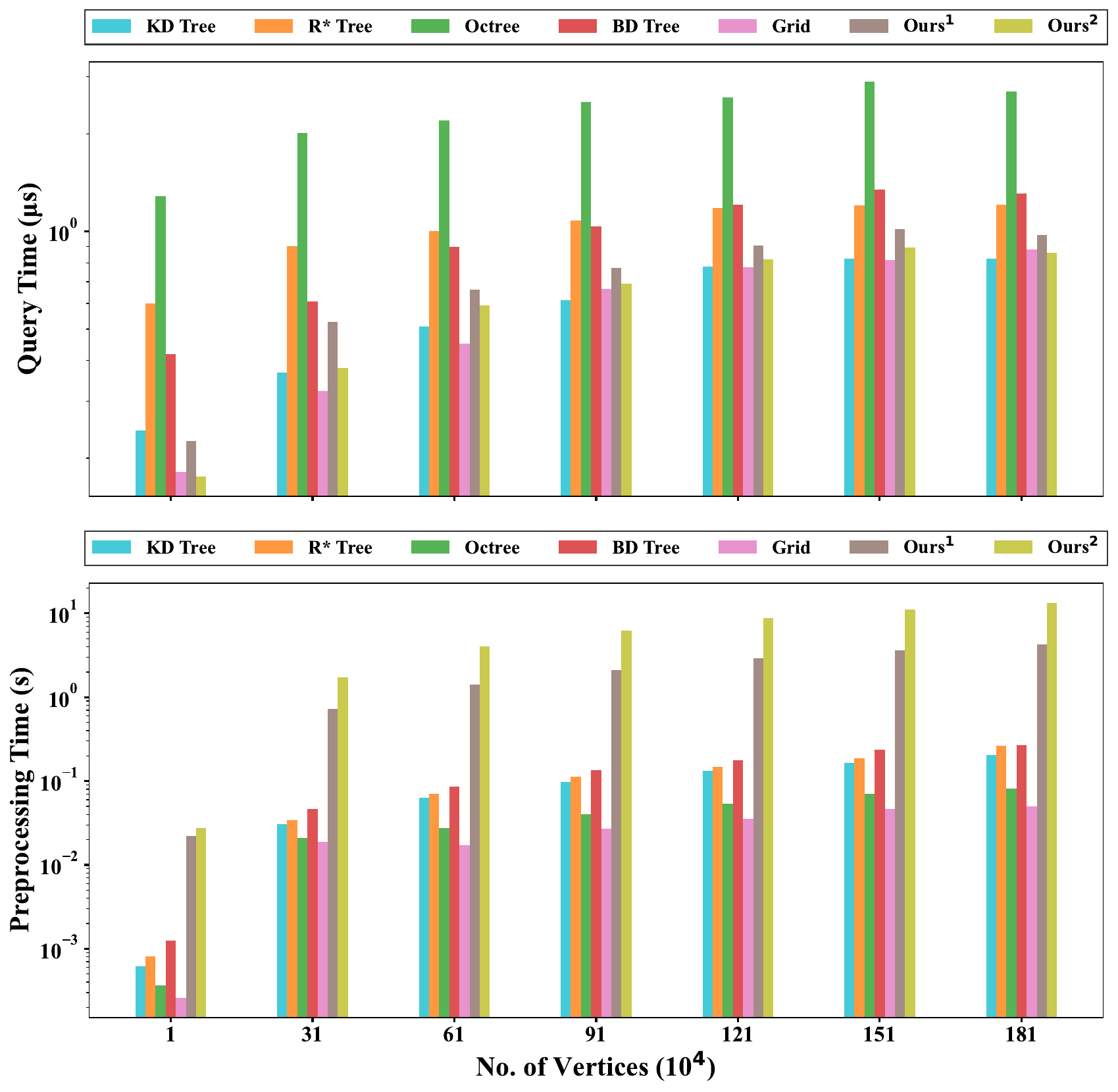}
\end{overpic}
\vspace{-3mm}
\caption{\CG{Performance evaluation on uniform point clouds with varying numbers of target points. Top: Query time comparison shows our method's efficiency comparable to KD Tree and Grid while outperforming others. Bottom: Preprocessing time comparison. 
All query points were generated within the exact same spatial bounds as the target points.}}
\label{Fig:RandomPointCloudNumVaryQuery0Scale}
\end{figure}

\CG{To evaluate our algorithm's efficiency across different data distributions, we tested performance on uniform point clouds. Unlike previous experiments, we generated query points within exactly the same spatial bounds as the target points—not in an expanded bounding box. As shown in Figure~\ref{Fig:RandomPointCloudNumVaryQuery0Scale}, which presents both query and preprocessing times, our method achieves query efficiency comparable to optimized KD Tree and Grid implementations while outperforming other approaches. These results demonstrate our algorithm's effectiveness across diverse point distribution patterns beyond its primary optimization for manifold data.}

\subsection{Near and Far Query Points.}
\begin{figure}[h]
	\hspace{-3mm} 
\begin{overpic}
[width=0.99\linewidth]{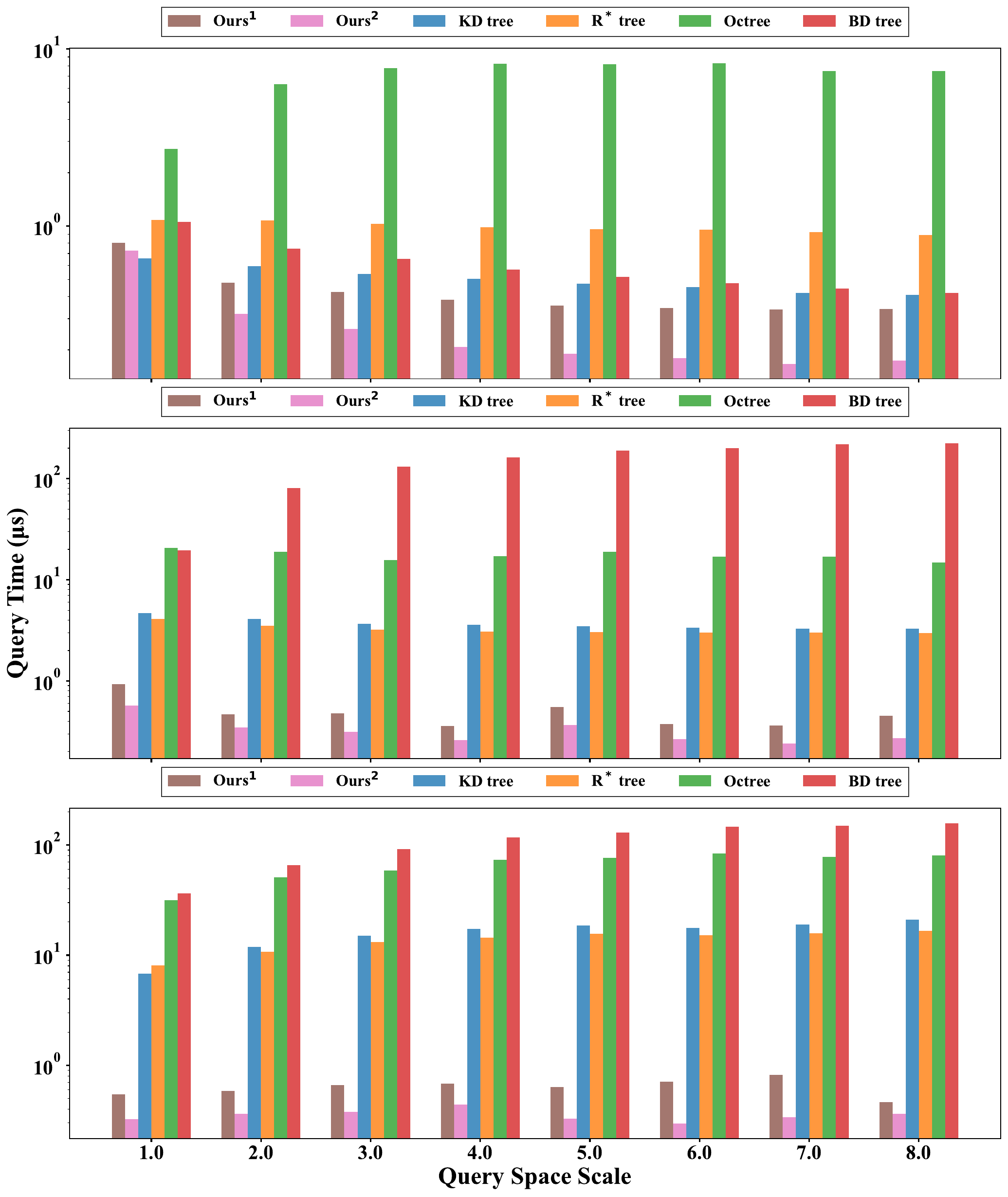}
\end{overpic}
\vspace{-3mm}
\caption{
\CG{Average query time with varying sampling box sizes. The horizontal axis shows the scaling factor applied to each dimension of the target point bounding box to define the query generation space. Results for uniform point cloud (top), Dragon model (middle), and Earth city data (bottom).}
}
\label{Fig:sizeofsamplebox}
\end{figure}

\CG{
The spatial relationship between query points and target points significantly influences nearest neighbor search efficiency. To quantify this effect, we conducted experiments varying the query space extent relative to the target point cloud. We tested three distinct distributions: uniform point clouds, Earth city models, and the Dragon dataset, each containing 1 million points.

Figure~\ref{Fig:sizeofsamplebox} presents query performance as a function of spatial scaling. The horizontal axis represents the ratio between query generation space and target point bounding box (e.g., a value of 2 indicates query points generated within a 2× axis-aligned bounding box). Corresponding preprocessing times are provided in Table~\ref{Table:BoxScaleTimeOnOtherTypeData}.

The results reveal an interesting pattern: even in uniform distributions, where our method shows comparable performance when query points share the same bounds as target points, our approach gains significant advantages as soon as the query space extends slightly beyond the target bounds.
For non-uniform distributions like Earth city models and Dragon dataset, our method consistently outperforms traditional approaches across all spatial configurations.
}

\begin{table}[]
\begin{center}
\caption{\CG{Preprocessing time comparison for uniform point cloud, Dragon model, and Earth city data (1M points each).}}
\label{Table:BoxScaleTimeOnOtherTypeData}
\vspace{-2mm}
\setlength{\tabcolsep}{0.1cm} 
\renewcommand{\arraystretch}{1.5} 
\resizebox{0.9\linewidth}{!}
{
\begin{tabular}{ccccccc}
\toprule
                             & Octree & BD Tree & R* Tree & KD Tree & Ours$^1$ & Ours$^2$ \\ \midrule
\multicolumn{1}{c|}{Uniform} & 0.047  & 0.145   & 0.121   & 0.105   & 2.317 & 6.775 \\
\multicolumn{1}{c|}{Dragon}  & 0.053  & 0.169   & 0.109   & 0.094   & 2.483 & 7.092 \\
\multicolumn{1}{c|}{Earth}   & 0.051  & 0.185   & 0.102   & 0.104   & 2.232 & 7.139 \\ \bottomrule
\end{tabular}
}
\end{center}
\end{table}

\section{Application}
\CG{
We validate the effectiveness of our algorithm through five applications.
}
\subsection{Point-to-Mesh Distance Queries}
\begin{figure}[h]
	\hspace{-3mm} 
\begin{overpic}
[width=0.99\linewidth]{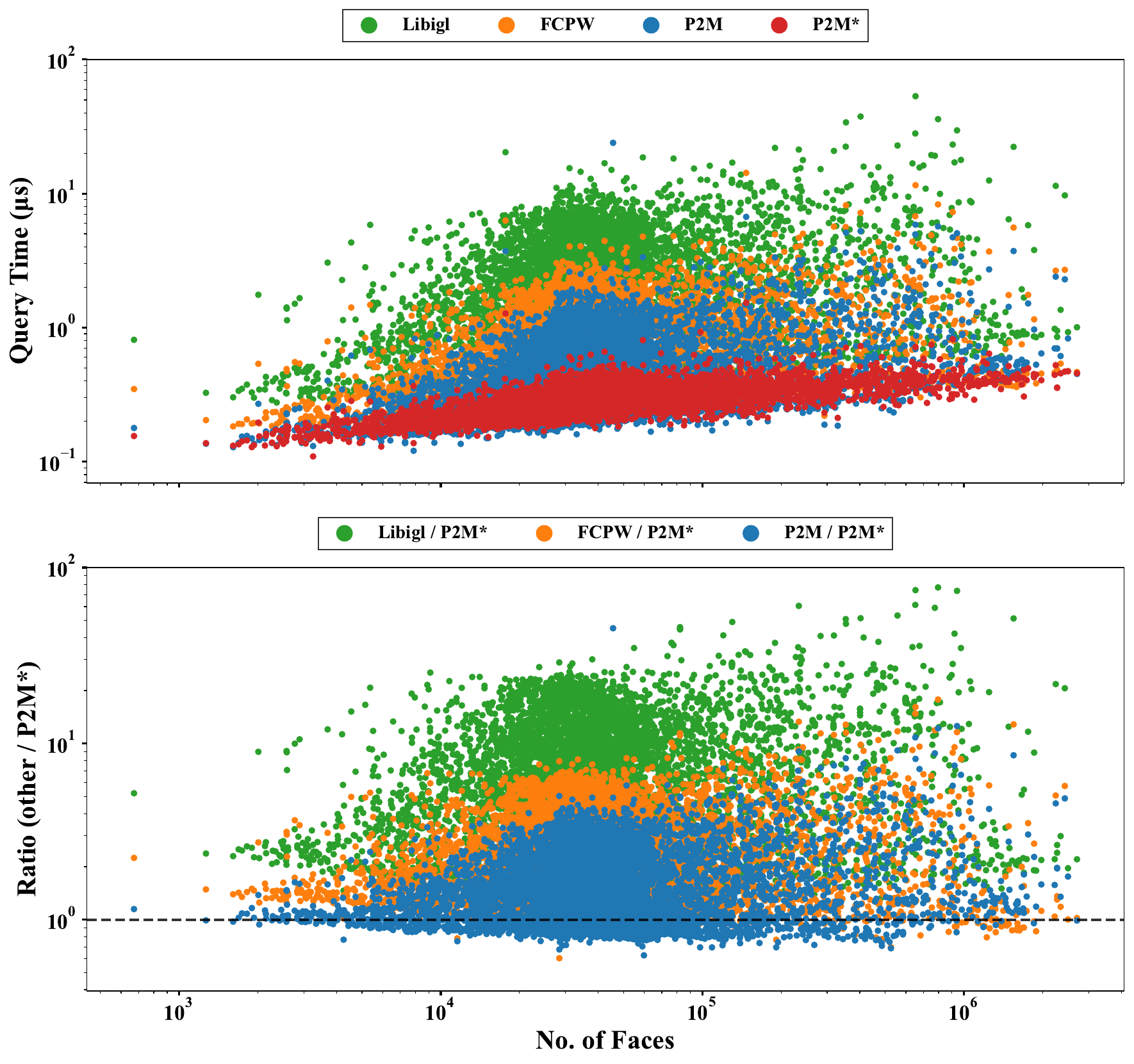}
\end{overpic}
\vspace{-3mm}
\caption{\CG{Query time comparison for point-to-mesh distance computation on the ABC dataset.}}
\label{Fig:Point2MeshABC}
\end{figure}

\CG{
Given a mesh surface, fast querying of the closest geometric primitive (vertex, edge, or face) to a user-specified point, along with the corresponding closest point and minimum distance, represents a fundamental operation across diverse research domains~\cite{Abbasifard,https://doi.org/10.1111/cgf.12228,910820,10.2312/hpg.20191189} including computer graphics, physical simulation, computational geometry, and computer-aided design. 
P2M~\cite{Zong2023P2M} currently stands as the fastest point-to-mesh distance query algorithm. During preprocessing, P2M constructs Voronoi diagrams of model point clouds and builds interception tables for each point cloud's Voronoi cell. In the query phase, it first employs a KD-tree to compute the nearest model vertex, then utilizes the interception table to obtain the nearest distance to the mesh.

In our experiments, we replace the KD-tree component with our algorithm. Since P2M already requires Voronoi diagram construction, our data structure introduces minimal additional preprocessing overhead while achieving \textbf{1-5× query speedup}.
We validate these speed improvements on the ABC dataset and several standard models, comparing against BVH-based implementations from Libigl~\cite{libigl} and FCPW~\cite{FCPW} libraries (with vectorization enabled), as well as the original P2M method. 
Following the experimental setup of the P2M paper, for each test model, we randomly sample a million query points within the 10× axis-aligned bounding box.
Query time results are presented in Figure~\ref{Fig:Point2MeshABC} and Table~\ref{Table:P2M_StandModel}. The modified P2M algorithm demonstrates significant efficiency gains, achieving 1-5× speedup over the original P2M query performance. Preprocessing times are shown in Figure~S1 and Table~S1 of the supplementary material, where our preprocessing overhead remains nearly identical to the original P2M implementation.
}

\begin{table}[]
\begin{center}
\caption{\CG{Query time comparison ($\mu$s) of different methods for point-to-mesh distance computation on standard models.
The \textbf{highest} and \underline{second}-fastest query times are highlighted in bold and underlined, respectively}}
\label{Table:P2M_StandModel}
\vspace{-2mm}
\setlength{\tabcolsep}{0.1cm} 
\renewcommand{\arraystretch}{1.5} 
\resizebox{1.0\linewidth}{!}
{
\begin{tabular}{cccccccc}
\toprule
                            & camel          & bunny          & armadillo      & dragon         & sponza         & kitten         & lucy           \\ \midrule
\multicolumn{1}{c|}{Faces}  & 19510          & 69451          & 99976          & 249882         & 262267         & 268896         & 525814         \\ \midrule
\multicolumn{1}{c|}{Libigl} & 3.104          & 7.032          & 5.521          & 3.897          & 0.813          & 17.03          & 4.350          \\
\multicolumn{1}{c|}{FCPW}   & 0.933          & 1.984          & 1.617          & 1.061          & \textbf{0.213} & 4.537          & 1.338          \\
\multicolumn{1}{c|}{P2M}    & {\ul 0.533}    & {\ul 1.457}    & {\ul 1.166}    & {\ul 0.854}    & 0.235          & {\ul 3.070}    & {\ul 0.883}    \\
\multicolumn{1}{c|}{P2M*}   & \textbf{0.243} & \textbf{0.367} & \textbf{0.344} & \textbf{0.342} & {\ul 0.229}    & \textbf{0.507} & \textbf{0.353} \\ \bottomrule
\end{tabular}
}
\end{center}
\end{table}

\subsection{Iterative Closest Point (ICP) Registration}
\begin{figure}[h]
	\hspace{-3mm} 
\begin{overpic}
[width=0.99\linewidth]{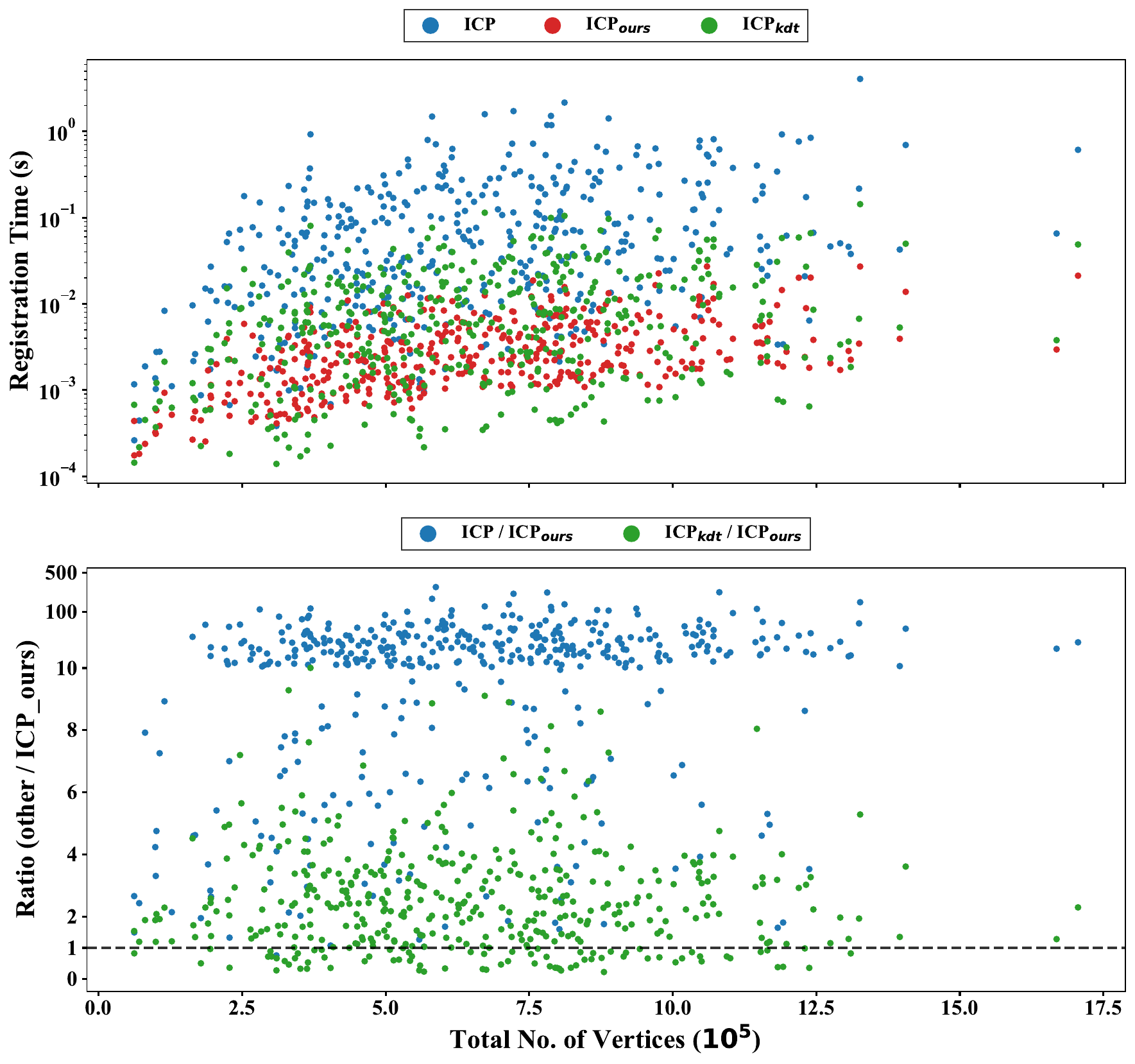}
\end{overpic}
\vspace{-3mm}
\caption{\CG{ICP registration speed comparison on 8 3DMatch test cases. 
The horizontal axis shows the sum of target and source point cloud sizes.
Our approach achieves faster performance on 82.4\% of the data.}}
\label{Fig:icpData}
\end{figure}

\CG{
Rigid registration, which finds an optimal rigid transformation to align a source point set with a target point set, is a fundamental problem in computer vision and many other areas. The Iterative Closest Point (ICP) algorithm~\cite{121791} is one of the most important algorithms for rigid registration. ICP operates iteratively, where in each iteration, every point in the source point set requires querying its nearest neighbor in the target point set, which significantly impacts the algorithm's efficiency.

In our experiments, we override the KD-tree implementation in PCL's~\cite{Rusu_ICRA2011_PCL} ICP algorithm, substituting it with our method and nanoflann's KD-tree implementation respectively. We evaluate three methods: the original ICP, ICP with our algorithm ($ICP_{ours}$), and ICP with nanoflann implementation ($ICP_{kdt}$) on 8 test scenarios from the 3DMatch dataset~\cite{zeng20163dmatch}, performing registration on all adjacent frame pairs. The timing results are shown in Figure~\ref{Fig:icpData} (statistics include complete runtime including both preprocessing and querying), where the x-axis represents the sum of points in the source and target point clouds.
Across all scenarios, our algorithm achieves faster registration in 82.4\% of the cases, with speedups exceeding 10× compared to $ICP_{kdt}$ in the best cases. Moreover, the advantage becomes more pronounced as the number of queries increases.
}

\begin{figure}[h]
	\hspace{-3mm} 
\begin{overpic}
[width=0.99\linewidth]{imgs/cluster.pdf}
\end{overpic}
\vspace{-3mm}
\caption{\CG{Four 2D clustering datasets from~\cite{10.1145/3448016.3452781,Birchsets}, each containing 1M points.}}
\label{Fig:ClusterExamples}
\end{figure}

\subsection{Density Peak Clustering in 2D}
\begin{figure}[h]
	\hspace{-3mm} 
\begin{overpic}
[width=0.99\linewidth]{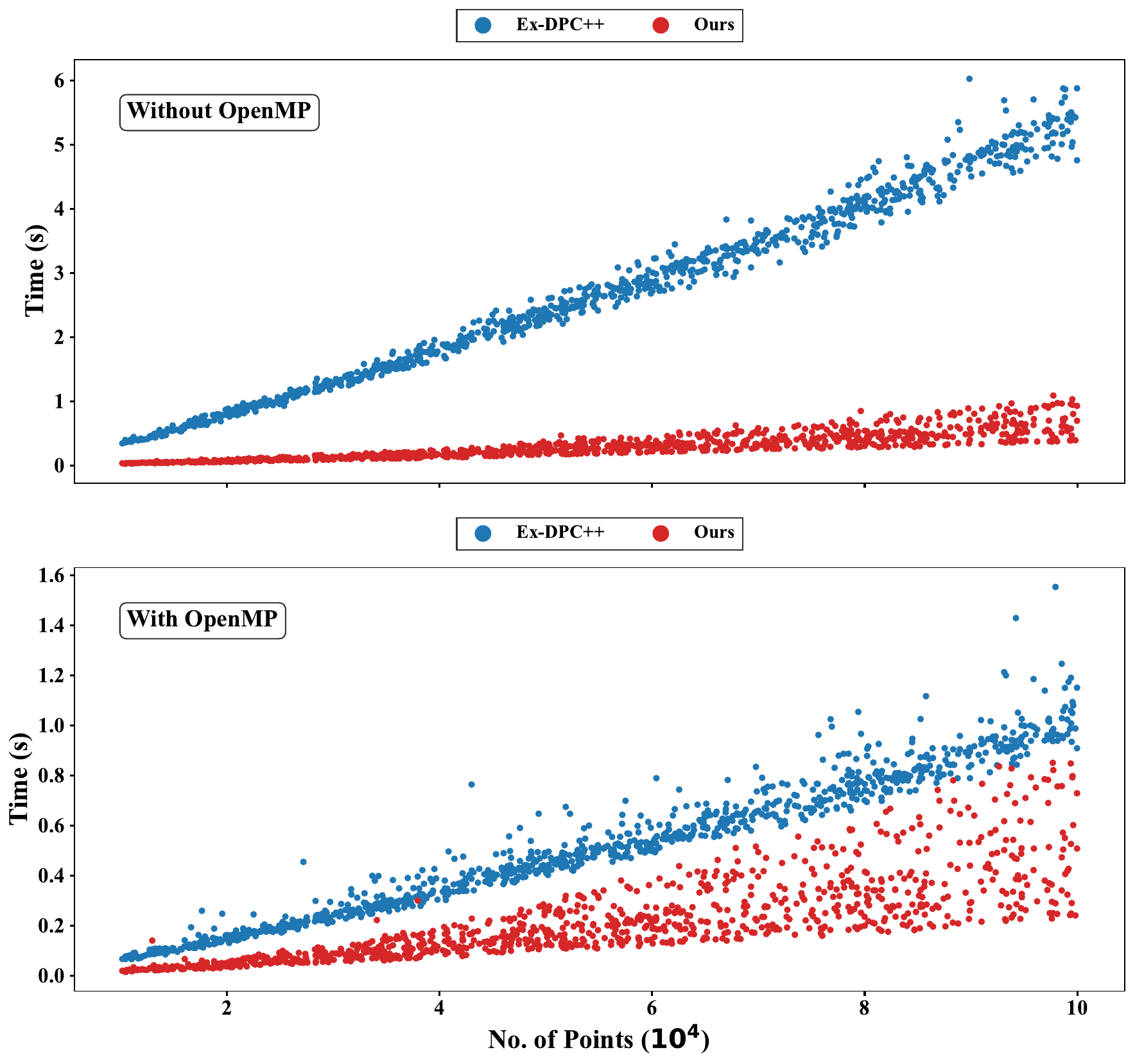}
\end{overpic}
\vspace{-3mm}
\caption{\CG{Clustering speed comparison on 1000 synthetic clustering data examples, without (up) and with (bottom) parallelization. Our method demonstrates consistent performance advantages in both scenarios.}}
\label{Fig:density_peaks_clustering}
\end{figure}

\CG{
Clustering is the task of grouping similar objects into clusters and represents a fundamental operation in data analysis and unsupervised machine learning. Clustering algorithms can be used to identify different types of tissues in medical imaging~\cite{YANG2002173},
analyze social networks, and identify weather regimes in climatology~\cite{ClusteringCoe}.
Among various clustering approaches, density-based clustering defines clusters as dense regions of points in the coordinate space and has received considerable attention. And Density peak clustering (DPC)~\cite{doi:10.1126/science.1242072} is a popular version
of density-based clustering.

A critical step in the DPC algorithm involves calculating $\delta_i$, which is defined as:
\begin{equation}
    \delta_i = \min_{j: \rho_j > \rho_i} (d_{ij}),
\end{equation}
where $\rho_i$ represents the density of point $p_i$, and $d_{ij}$ denotes the distance between points $p_i$ and $p_j$. Generally, computing $\{\delta_i\}_{i=1}^n$ requires calculating distances between every pair of points, resulting in $O(n^2)$ complexity. Numerous algorithms have been proposed to accelerate the computation of $\delta_i$ through heuristic approaches~\cite{amagata2024dpc,10.1145/3597635.3598021}.

One notable property of our algorithm is its ability to efficiently identify the nearest point among $\{p_i\}_{i=1}^{k}, k \leq n$ for any query point $q$. 
As discussed in Section~\ref{Sec:Methodology}, during the nearest-neighbor search process, when querying the nearest neighbor, we only need to compute \(\Phi_{n}(q)\). However, in practice, \(\forall~i \in [1,n]\), \(\Phi_{i}(q)\) is already incorporated into our query process. When a specific $k$ is provided, as shown in Algorithm~\ref{Algorithm:Query}, our query process can terminate early. To the best of our knowledge, no existing algorithm achieves this capability.

Therefore, in 2D spaces, our method is particularly well-suited for accelerating density peak clustering algorithms. We simply sort the point cloud by density and use this ordering for incremental preprocessing in our algorithm. In our experiments, we replace the $\delta$ computation process in Ex-DPC++~\cite{amagata2024dpc}, the fastest known density peak clustering algorithm as we known, with our method. We evaluate performance on four classical scenarios~\cite{10.1145/3448016.3452781,Birchsets}, each containing 100,000 points, as shown in Figure~\ref{Fig:ClusterExamples}, and 1000 randomly generated 2D clustering scenarios. Table~\ref{Table:clusterTime} and Figure~\ref{Fig:density_peaks_clustering} present the comparative results. Without parallelization, our clustering achieves over 10× speedup compared to the original Ex-DPC++. Even with parallelization enabled (while our $\delta$ computation remains serial), our method maintains 1-3× speed advantages.
It should be noted that due to the complexity of high-dimensional Delaunay construction, our strategy-accelerated density peak clustering algorithm is currently suitable only for low-dimensional spaces (2D/3D), with high-dimensional clustering requiring further investigation.

Beyond the speed advantages, our method offers stronger theoretical foundations compared to existing heuristic-based and parallelized algorithms. Additionally, our capability to query the nearest point among the first $k$ points may find applications in various practical scenarios:

\begin{enumerate}
    \item Finding the most similar weather conditions to current patterns that occurred before a specific date for climate analysis.
    \item Identifying the closest historical stock price pattern to current market conditions, restricted to periods before specific regulatory changes.
    \item In mobile location services, users seek restaurants that are both ``affordable and nearby." The system needs to identify the nearest restaurant among all lower-priced establishments..
    \item Finding ``highly qualified and role-appropriate" benchmark employees for job candidates during recruitment processes.
    \item Journal editors need to identify reviewers who are both ``academically distinguished and research-field relevant" for submitted manuscripts.
\end{enumerate}
}

\begin{table}[]
\begin{center}
\caption{\CG{Clustering speed comparison (in seconds) on four clustering data examples, before and after OpenMP parallelization. Our method demonstrates consistent speed advantages across all examples.}}
\label{Table:clusterTime}
\vspace{-2mm}
\setlength{\tabcolsep}{0.1cm} 
\renewcommand{\arraystretch}{1.5} 
\resizebox{0.8\linewidth}{!}
{
\begin{tabular}{lllll}
\toprule
                                & syn-2d & Birch1 & Birch2 & Birch3 \\ \midrule
\multicolumn{1}{l|}{Ours}       & 0.715  & 0.432  & 0.577  & 0.796  \\ \cline{1-1}
\multicolumn{1}{l|}{Ex-DPC++}   & 4.974   & 5.723   & 4.961   & 5.386   \\ \midrule
\multicolumn{1}{l|}{Ours (omp)} & 0.559  & 0.298  & 0.407  & 0.610  \\ \cline{1-1}
\multicolumn{1}{l|}{Ex-DPC++ (omp)}   & 0.927  & 0.953  & 0.1008 & 0.960  \\ \bottomrule
\end{tabular}
}

\end{center}
\end{table}

\subsection{Point-to-Segments Distance Query}
\begin{figure}[h]
	\hspace{-3mm} 
\begin{overpic}
[width=0.99\linewidth]{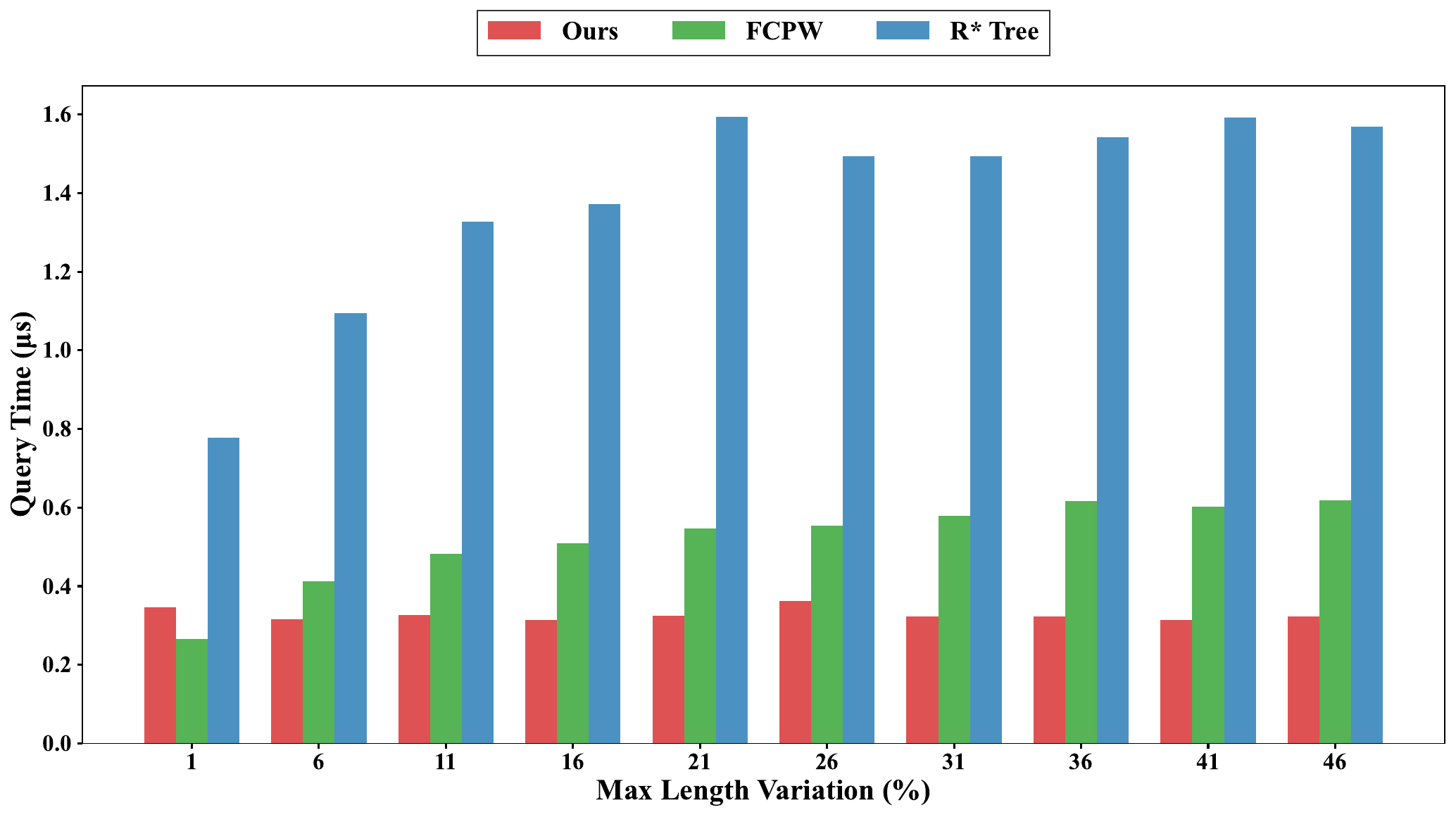}
\end{overpic}
\vspace{-3mm}
\caption{\CG{Point-to-segments distance query performance comparison with varying maximum segment lengths.}}
\label{Fig:PointToSegmentsDistanceQuery}
\end{figure}

\CG{In 2D scenarios, computing the nearest distance from a point to a collection of line segments represents a fundamental task in computer graphics, with widespread applications in path planning, navigation, and collision detection. Traditional approaches for nearest segment queries typically employ hierarchical bounding volume strategies, such as bounding volume hierarchies (BVH) or R-trees, to efficiently organize and traverse the line segment collection.

Although this paper primarily focuses on computing nearest distance queries from points to point sets on 2D manifold surfaces in 3D space, our nearest neighbor search framework can be extended to different primitives, dimensions, distance metrics, and even solution spaces, provided that incremental Delaunay construction algorithms exist for the specific scenario and a distance query function from any point in the computational space to the primitives is available. Leveraging CGAL's integrated 2D segment Delaunay construction algorithm~\cite{segmentVoronoi}, our nearest neighbor search approach can be adapted for point-to-disjoint-segments nearest distance queries in 2D.

In our experiments, we randomly generate 50,000 non-intersecting line segments within the domain $[-5,5]^2$, with segment lengths randomly distributed in the range $[0,\delta]$. By controlling the parameter $\delta$, we obtain multiple test scenarios with varying geometric complexity. We randomly generate 1 million query points within $[-5,5]^2$ and compare against FCPW and R*Tree (from Boost). Figure~\ref{Fig:PointToSegmentsDistanceQuery} presents the comparative results, where the x-axis represents the percentage of $\delta$ relative to the diagonal length of the generation space ($5\sqrt{2}$), and the y-axis shows the average query time per operation. As observed, both R*Tree and BVH-based methods exhibit declining query efficiency as the maximum segment length increases. In contrast, our method maintains consistent query efficiency across scenarios with different maximum segment lengths while demonstrating superior performance compared to existing approaches.
Preprocessing times are shown in Figure~S2 of the supplementary material.
}

\subsection{Farthest Point Sampling as a Natural Extension}
\CG{
Although our method was not specifically designed for farthest point sampling, our incremental Voronoi construction framework naturally extends to support this functionality. The algorithm can readily identify farthest points during the incremental construction process, making it a straightforward derivation of our core approach.

While our naturally derived implementation is slower than existing specialized farthest point sampling algorithms~\cite{10122654}, it demonstrates the versatility of our framework and provides a simple solution that emerges directly from our Voronoi construction methodology. The comparative performance results are presented in Table~\ref{table:farthestPointSampling}.}


\begin{table}[]
\begin{center}
\caption{\CG{Farthest point sampling time comparison between QuickFPS and our method at 25\% sampling rate on standard models.}}
\label{table:farthestPointSampling}
\vspace{-2mm}
\setlength{\tabcolsep}{0.1cm} 
\renewcommand{\arraystretch}{1.5} 
\resizebox{1\linewidth}{!}
{
\begin{tabular}{cccccccc}
\toprule
\multicolumn{1}{l}{}          & Camel & Bunny & Dragon & Kitten & Armadillo & Lucy    & Sponza  \\ \midrule
\multicolumn{1}{c|}{Vertices} & 28934 & 72911 & 100313 & 291023 & 726367    & 1018219 & 1313504 \\ \midrule
\multicolumn{1}{c|}{QuickFPS} & 0.023 & 0.064 & 0.087  & 0.265  & 0.865     & 1.401   & 2.013   \\ \cline{1-1}
\multicolumn{1}{c|}{Ours}     & 0.035 & 0.102 & 0.142  & 0.610  & 2.361     & 3.428   & 4.501   \\ \bottomrule
\end{tabular}
}
\end{center}
\end{table}

\section{Limitations and Future Work}
\label{Sec:limitationsAndFutureWork}
\CG{The primary limitation of our algorithm is its high preprocessing cost and the difficulty of parallelizing the preprocessing phase,}
which restricts its applicability, particularly in large-scale scenarios or with high-dimensional datasets where efficiency is crucial. Therefore, in future work, we plan to couple the construction of our algorithm more closely with Delaunay computations, or alternatively, develop an approximate Delaunay construction to achieve faster preprocessing. 
\CG{Additionally, we will explore the potential for accelerating the preprocessing phase through parallelization strategies. These objectives aim to reduce the preprocessing burden and improve scalability for large-scale and high-dimensional applications.}

Additionally, our method currently only supports nearest neighbor search. In future work, we aim to extend the algorithm to support k-nearest neighbor queries, which would enhance its versatility. 
\CG{Furthermore, our approach is primarily effective in low-dimensional spaces (2D/3D), as high-dimensional Delaunay triangulation becomes computationally expensive with prohibitive preprocessing costs. Exploring approximate Delaunay construction techniques for high-dimensional spaces represents another promising direction that could potentially overcome this dimensional constraint while maintaining reasonable preprocessing costs.}

\section{Conclusion}

In this paper, we observe that traditional nearest neighbor search algorithms exhibit decreased efficiency when dealing with point clouds distributed over 2D manifolds in 3D space, or when query points are distant from the target points. In certain special cases, the complexity can degrade to \(O(n)\).
\CG{
Inspired by the incremental Delaunay construction process, we propose a novel nearest neighbor search algorithm specifically designed for 2D manifold point cloud data. Our method achieves 1-10× speedup compared to KD-Tree and R*-Tree approaches, with efficiency advantages validated through extensive experiments.
}


\CG{
Additionally, our algorithm demonstrates significant practical potential. We validated the effectiveness and versatility of our approach through five diverse applications: point-to-mesh distance queries, ICP registration, 2D density peak clustering, point-to-segments distance queries, and farthest point sampling.
We believe that the algorithm presented in this paper offers substantial practical value and theoretical contributions that will benefit the broader research community.
}





\bibliographystyle{IEEEtran}
\bibliography{main}



 

\begin{IEEEbiography}[{\includegraphics[width=1in,height=1.25in,clip,keepaspectratio]{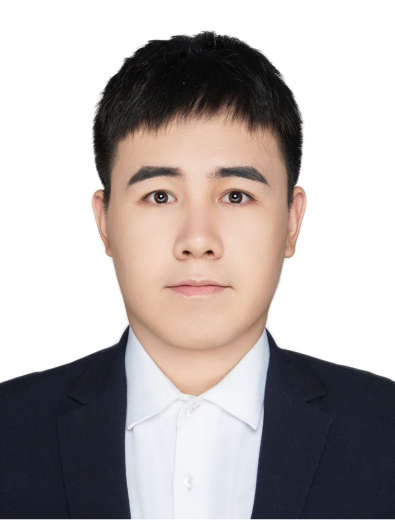}}]{Pengfei Wang}
is currently working toward the PhD degree in computer science with Shandong University. His research interests include computational geometry, computer graphics, and computer-aided design.
\end{IEEEbiography}

\begin{IEEEbiography}[{\includegraphics[width=1in,height=1.25in,clip,keepaspectratio]{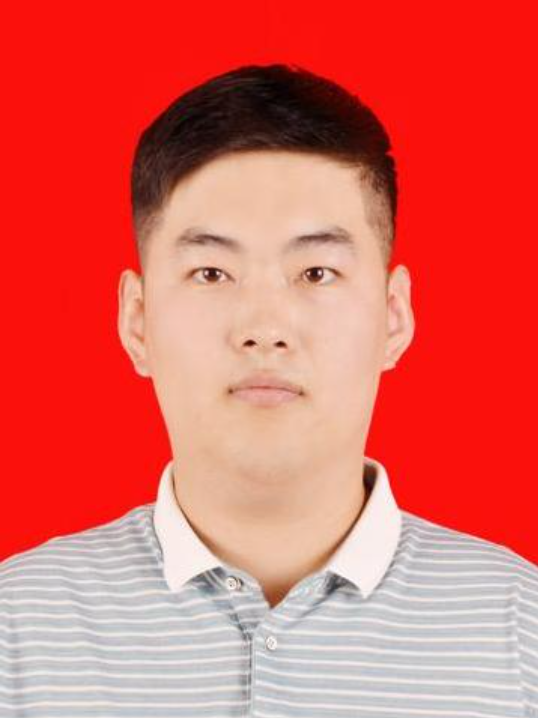}}]{Jiantao Song}
is currently pursuing a Ph.D. in computer science at Shandong University His research interests are computer graphics, computational geometry, point cloud reconstruction and geometric modeling.
\end{IEEEbiography}

\begin{IEEEbiography}[{\includegraphics[width=1in,height=1.25in,clip,keepaspectratio]{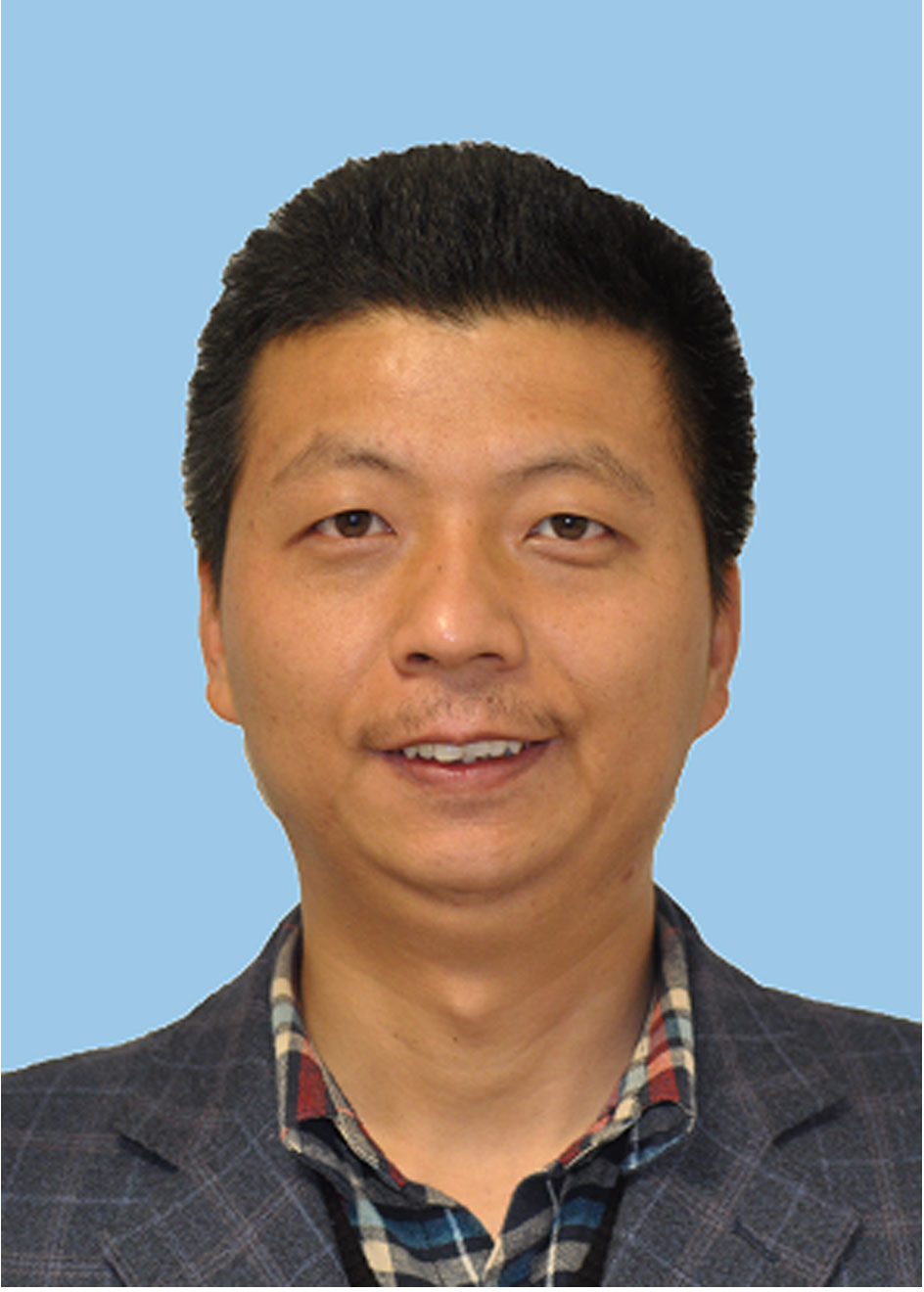}}]{Shiqing Xin}
is an associate professor at the School of Computer Science and Technology, Shandong University. He received his Ph.D. degree in applied mathematics from Zhejiang University. His research focuses on geometric calculation, geometric modeling, and scene understanding.
\end{IEEEbiography}

\begin{IEEEbiography}[{\includegraphics[width=1in,height=1.25in,clip,keepaspectratio]{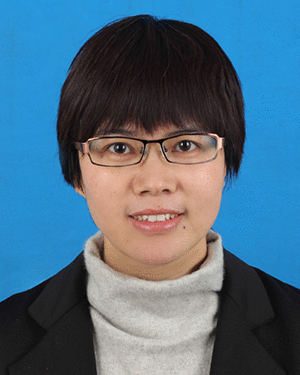}}]{Shuangmin Chen}
received the PhD degree from Ningbo University in 2018. She is currently an assistant professor with the School of Information and Technology, Qingdao University of Science and Technology, China. She has authored or coauthored more than 40 research papers in famous journals and conferences. Her research interests include computer graphics and computational geometry
\end{IEEEbiography}

\begin{IEEEbiography}[{\includegraphics[width=1in,height=1.25in,clip,keepaspectratio]{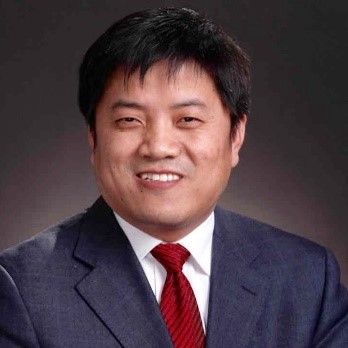}}]{Changhe Tu}
received the BSc, MEng, and PhD degrees from Shandong University, China, in 1990, 1993, and 2003, respectively. He is currently a professor with the School of Computer Science and Technology, Shandong University, China. He currently leads the CG-VIS Group, Shandong University. He has authored or coauthored more than 100 papers in international journals and conferences. His research interests include computer graphics, 3D vision, and computer-aided geometric design.
\end{IEEEbiography}

\begin{IEEEbiography}[{\includegraphics[width=1in,height=1.25in,clip,keepaspectratio]{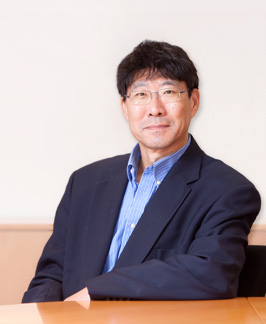}}]{Wenping Wang}
(Fellow, IEEE) received the PhD degree in computer science from the University of Alberta, in 1992. He is currently a chair professor of Computer Science with the University of Hong Kong. His research interests include computer graphics, computer visualization, computer vision, robotics, medical image processing, and geometric computing. He is associate editor of several premium journals, including the Computer Aided Geometric Design (CAGD), Computer Graphics Forum (CGF), IEEE Transactions on Computers, and IEEE Computer Graphics and Applications, and has chaired more than 20 international conferences, including Pacific Graphics 2012, ACM Symposium on Physical and Solid Modeling (SPM) 2013, SIGGRAPH Asia 2013, and Geometry Submit 2019. He received the John Gregory Memorial Award for his contributions in geometric modeling.
\end{IEEEbiography}

\begin{IEEEbiography}[{\includegraphics[width=1in,height=1.25in,clip,keepaspectratio]{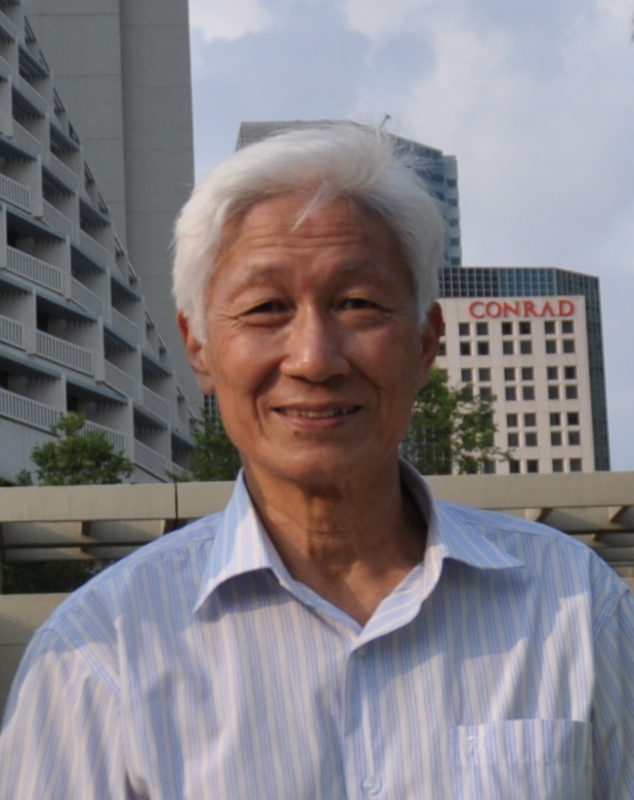}}]{Jiaye Wang}
graduated from the Dept. of Mathematics of Shandong University in 1959. Now he is a Professor of Dept. of Computer Science and Technology of Shandong University. He has gone to the University of East Anglia, UK as an invited visitor in 1979, and has been a  Senior Fellow at GINTIC Institute of Manufacturing Technology of Singapore from 1991. Prof. Wang has led several national, provincial and departmental projects, and published about 50 papers in domestic and foreign major publications and conferences in the related areas.
\end{IEEEbiography}



\end{document}